\documentclass[showpacs, pra,onecolumn,preprintnumbers ,amsmath, amssymb, superscriptaddress, aps]{revtex4-2}
\usepackage{color}
\usepackage{amsmath,amssymb}
\usepackage{pifont}
\usepackage{amssymb}  
\usepackage{bbold}
\usepackage{float}
\usepackage{subfloat}
\usepackage[squaren]{SIunits}

\usepackage[caption=false]{subfig}
\usepackage{tikz}
\usepackage{makecell}
\usepackage{subfig}
\usepackage{pifont}   
\usepackage{graphicx} 
\graphicspath{{Figures4/}}
\usepackage{dcolumn}  
\usepackage{bm}       
\usepackage{multirow} 
\usepackage{placeins}
\usepackage[colorlinks]{hyperref}
\usepackage{mathtools}
\usepackage{appendix}

\newcommand{\tr}{\mathrm{tr}}

\begin{document}
	
	\title{Entanglement distribution in pure non-Gaussian tripartite states: a Schmidt decomposition approach}
	\date{\today}
	
	\author{Abdeldjalil Merdaci}
	\email{a.merdaci@univ-skikda.dz}
	\affiliation{Department of Physics, Faculty of Sciences
		University of 20 Août 1955-Skikda
		Road El-Hadaeik, B.P. 26, 21000 Skikda, Algeria}
	\author{Ahmed Jellal}
	\email{a.jellal@ucd.ac.ma}
	\affiliation{Laboratory of Theoretical Physics, Faculty of Sciences, Choua\"ib Doukkali University, PO Box 20, 24000 El Jadida, Morocco}
	\affiliation{Canadian Quantum  Research Center,
		204-3002 32 Ave Vernon,  BC V1T 2L7,  Canada}

	\begin{abstract}

	We study entanglement in a system of three coupled quantum harmonic oscillators. Specifically, we use the Schmidt decomposition to analyze how the entanglement is distributed among the three subsystems. The Schmidt decomposition is a powerful mathematical tool for characterizing bipartite entanglement in composite quantum systems. It allows to write a multipartite quantum state as a sum of product states between the subsystems, with coefficients known as Schmidt coefficients.
	We apply this decomposition to the general quantum state of three coupled oscillators and study how the Schmidt coefficients evolve as the interaction strengths between the oscillators are varied. This provides insight into how entanglement is shared between the different bipartitions of the overall three-particle system. Our results advance the fundamental understanding of multipartite entanglement in networked quantum systems. They also have implications for quantum information processing using multiple entangled nodes.

\end{abstract}


\maketitle

\section{Introduction}

Entanglement is a uniquely quantum mechanical phenomenon with no classical analog that lies at the heart of many phenomena in physics. It has been extensively studied in the context of bipartite systems, i.e. systems divided into two parts \cite{1,2,3,4,5}. However, multipartite entanglement, which involves more than two entangled parties, exhibits a richer structure and is crucial for applications such as quantum secret sharing \cite{06,07}.  There have been many investigations of entanglement properties in bipartite settings, including coupled harmonic oscillators \cite{08,09,010}, spin systems \cite{11,12}, and itinerant particle models \cite{013,014}. For a pure state of two parties $A$ and $B$, the Schmidt decomposition \cite{015} provides the most useful quantification and characterization of entanglement by writing the state as a unique superposition of product states involving the $A$ and $B$ subsystems. The corresponding Schmidt coefficients quantify the amount of bipartite entanglement. 

Makarov studied the solutions of coupled harmonic oscillators and their quantum entanglement. He derived a solution to the nonstationary Schrödinger equation and provided an analytical solution for the Schmidt modes in both the stationary and dynamic cases. Using these Schmidt modes, he analyzed the quantum entanglement of the system and showed that under certain system parameters the quantum entanglement can become extremely large.
Multipartite entanglement is much more complex due to the larger number of partitions and the possibility of quantum correlations. For three parties $A, B,$ and $C$ in a pure state, the natural generalization of the Schmidt decomposition expresses the global state as a sum over product states involving all bipartitions \cite{016,017}. However, an infinite number of Schmidt coefficients can be nonzero in general, as opposed to the maximally two nonzero coefficients for bipartite states. This richer structure requires novel tools and measures for the classification and detection of tripartite and higher multipartite entanglement \cite{018,019,020}.
Experimental platforms that can realize three-body entanglement include trapped ion crystals \cite{021,022,023}, superconducting qubits \cite{024,025}, cavity and circuit QED \cite{026,027,028}, and optomechanical systems \cite{029,030,031}. However, a detailed understanding of how specific model Hamiltonians generate different types of multipartite entanglement remains an important open question.

Makarov's approach \cite{Makarov} to studying the entanglement of two harmonic oscillators using Schmidt decomposition was restricted to a single parameter, namely the mixing angle. In contrast, our analysis will focus solely on entanglement as a function of the mixing angles. More precisely, 
we study tripartite entanglement generated in the simplest setting of three coupled quantum harmonic oscillators. Using the tripartite Schmidt decomposition, we analyze the entanglement properties of the system in different parameter regimes. We derive analytic solutions in certain limits to provide physical insight. Our work contributes to the characterization of the subtleties of three-body entanglement and may aid in the engineering of tripartite entangled states for quantum technologies.

The present paper is organized as follows. In Sec. \ref{MSD2}, we present our theoretical model, which is based on three coupled quantum harmonic oscillators, along with the solutions for the energy spectrum. We then focus on the tripartite Schmidt decomposition and show how the eigenstates allow us to derive the Schmidt coefficients necessary for the analysis of entangled states. In Sec. \ref{Entag}, we present the main contribution of our work: the use of the Schmidt coefficients to study the entanglement within the system. To illustrate this, we provide several examples and numerical discussions. Finally, we summarize our findings in the conclusion.

\section{Model and Schmidt decomposition}\label{MSD2}

We start by considering the general quantum state of the three oscillators under a Hamiltonian with coupling between them. Provided that the coupling strengths are sufficiently small, the energy eigenstates take a separable form in terms of harmonic oscillator wavefunctions for each particle. 
In our previous work, we studied the entanglement of three coupled harmonic oscillators based on the algebraic structure of the Lie group $SU(3)$ \cite{MerdaciPLA2020}. More specifically, we considered the following Hamiltonian  
\begin{equation}\label{Ham1}
H=\frac{1}{2}\left(  \frac{p_{1}^{2}}{m_{1}}+\frac{p_{2}^{2}}{m_{2}}%
+\frac{p_{3}^{2}}{m_{3}}+m_{1}\omega_{1}^{2}x_{1}^{2}+m_{2}\omega_{2}^{2}%
x_{2}^{2}+m_{3}\omega_{3}^{2}x_{3}^{2}+D_{1}x_{1}x_{2}+D_{2}x_{1}x_{3}%
+D_{3}x_{2}x_{3}\right)
\end{equation}
and showed that the corresponding eigenstates are
 the following wavefunctions written in terms of the Hermite polynomials (see Appendix \ref{EESS})
\begin{align}\label{statess}
\psi_{n_{1},n_{2},n_{3}}^{ABC}\left(  x_{1},x_{2},x_{3}\right)    =&\left(
\frac{m\varpi}{\pi\hbar}\right)  ^{\frac{3}{4}}\frac{1}{\sqrt{2^{n_{1}%
+n_{2}+n_{3}}n_{1}!n_{2}!n_{3}!}}e^{-\tfrac{m\varpi}{2\hbar}\left(
e^{\varsigma-\rho}q_{1}^{2}+e^{\kappa-\varsigma}q_{2}^{2}+e^{\rho-\kappa}%
q_{3}^{2}\right)  }\nonumber\\
&   H_{n_{1}}^A\left(  \sqrt{\tfrac{m\varpi e^{\varsigma-\rho}}{\hbar}%
}q_{1}\right)  H_{n_{2}}^B\left(  \sqrt{\tfrac{m\varpi e^{\kappa-\varsigma}%
}{\hbar}}q_{2}\right)  H_{n_{3}}^C\left(  \sqrt{\tfrac{m\varpi e^{\rho-\kappa}%
}{\hbar}}q_{3}\right)
\end{align}
where $A$, $B$, and $C$ refer to the first, second, and third harmonic oscillators, respectively. These wavefunctions will serve as the basis for explaining an alternative approach to exploring entanglement in our system.

For the upcoming analysis, our goal is to investigate the quantum entanglement of three oscillators using Schmidt modes, focusing on a particular scenario. This requires that the coupling parameters $J_{ij}$, defined by (\ref{j12}-\ref{j23}) in Appendix \ref{EESS}, are significantly small ($J_{ij}\ll1$). This requirement can be satisfied by configuring the system so that
\begin{equation}
	J_{ij}=\epsilon_{ij}\tilde{J}_{ij},\quad\epsilon_{ij}\sim0
	\end{equation}
and at the same time we take the oscillator frequencies $\omega_{i}$ (\ref{o22}-\ref{o33}) similar to each other with the same order as $J_{ij}$, then we write
\begin{equation}
\omega_{i}^{2}-\omega_{j}^{2}=\epsilon_{ij}\left(  \tilde{\omega}_{i}%
^{2}-\tilde{\omega}_{j}^{2}\right).
\end{equation}
Under the assumption $\varsigma\sim\rho\sim\kappa$, that is, $\Sigma_{1}^{2}\sim
\Sigma_{2}^{2}\sim\Sigma_{3}^{2}$, we can write
\begin{equation}
\Sigma_{1}^{2}-\Sigma_{2}^{2}=\Sigma_{2}^{2}-\Sigma_{3}^{2}=\frac{1}{2}\left(
\Sigma_{1}^{2}-\Sigma_{3}^{2}\right)  =\varepsilon,\quad\varepsilon\sim0
\end{equation}
and it follows that (\ref{A14}-\ref{A16}) can be reduced to 
\begin{align}
&  \tfrac{2\tilde
{J}_{12}}{\tilde{\omega}_{1}^{2}-\tilde{\omega}_{2}^{2}}=\tfrac{-\cos
2\theta\sin2\phi\cos\varphi-2\sin2\theta\cos\phi\sin\varphi}{\sin^{2}\phi
-\cos^{2}\phi\cos^{2}\varphi-2\sin^{2}\theta\sin^{2}\varphi+2\cos^{2}%
\theta\cos^{2}\phi-2\cos^{2}\theta\sin^{2}\phi\cos^{2}\varphi-\sin2\theta
\sin\phi\sin2\varphi}\label{200}\\
&  \tfrac{2\tilde
{J}_{13}}{\tilde{\omega}_{1}^{2}-\tilde{\omega}_{3}^{2}}=\tfrac{\cos
2\theta\sin2\phi\sin\varphi-2\sin2\theta\cos\phi\cos\varphi}{-2\sin^{2}%
\theta\cos^{2}\varphi+2\cos^{2}\theta\cos^{2}\phi-2\cos^{2}\theta\sin^{2}%
\phi\sin^{2}\varphi+\sin2\theta\sin\phi\sin2\varphi+\sin^{2}\phi-\cos^{2}%
\phi\sin^{2}\varphi}\\
& \tfrac{2\tilde
{J}_{23}}{\tilde{\omega}_{2}^{2}-\tilde{\omega}_{3}^{2}}=\tfrac{-2\cos
2\theta\sin^{2}\phi\sin2\varphi+2\sin2\theta\sin\phi\cos2\varphi-\cos
2\theta\cos^{2}\phi\sin2\varphi}{\cos^{2}\phi\cos2\varphi-2\sin^{2}\theta
\cos2\varphi+2\cos^{2}\theta\sin^{2}\phi\cos2\varphi+2\sin2\theta\sin\phi
\sin2\varphi}\label{211}.
\end{align}
These relations make it clear that the new parameters $\tilde{J}_{ij}$ and $\tilde{\omega}_{i}$, if they span their entire range of values, will encompass the entire set of real numbers. Consequently, the tangents $\left(\tan 2\theta ,\tan 2\varphi ,\tan 2\phi \right)$ are not subject to any restrictions, and then we can assume values such that $\theta, \varphi, \phi\in\left[ -\frac{\pi}{4},+\frac{\pi}{4}\right]$.
Considering all the above conditions, we can express the wavefunctions \eqref{statess} of the system as follows
\begin{align}\label{abc}
\psi_{n_{1},n_{2},n_{3}}^{ABC}\left(  x_{1},x_{2},x_{3}\right)    =\tfrac{ \left(
	\tfrac{m\varpi}{\pi\hbar}\right)  ^{\frac{3}{4}}}{\sqrt{2^{n_{1}%
+n_{2}+n_{3}}n_{1}!n_{2}!n_{3}!}}e^{-\tfrac{m\varpi}{2\hbar}\left(  q_{1}%
^{2}+q_{2}^{2}+q_{3}^{2}\right)  } H_{n_{1}}^A\left(  \sqrt{\tfrac{m\varpi}{\hbar}}q_{1}\right)
H_{n_{2}}^B\left(  \sqrt{\tfrac{m\varpi}{\hbar}}q_{2}\right)  H_{n_{3}}^C\left(
\sqrt{\tfrac{m\varpi}{\hbar}}q_{3}\right)
\end{align}
and the associated eigenvalues
\begin{equation}
E_{n_{1},n_{2},n_{3}}=\hbar\varpi\left(  n_{1}+n_{2}+n_{3}+\frac{3}{2}\right).
\end{equation}
At this stage, we have established all the essential tools required to study the Schmidt decomposition of the wavefunction needed to study the entanglement of for three coupled harmonic oscillators.

Before going any further, we emphasize that the groundstate wavefunction is desentangled because we have separability in the zero mode. Now we decompose the wavefunction \eqref{abc} as
\begin{equation}
\psi_{n_{1},n_{2},n_{3}}^{ABC}\left(  x_{1},x_{2},x_{3}\right)  =\sum
_{l,k,m=0}^{\infty}A_{n_{1},n_{2},n_{3}}^{k,l,m}\ \varphi_{k}^A\left(
x_{1}\right)\  \phi_{l}^B\left(  x_{2}\right)\  \chi_{m}^C\left(  x_{3}\right)
\label{244}%
\end{equation}
where $\varphi_{k}^A\left(x_{1}\right)$, $\phi_{l}^B\left(x_{2}\right)$, and $\chi_{m}^C\left(x_{3}\right)$ take the forms
\begin{align}
&\varphi_{k}^A\left(  x_{1}\right)     =\left(  \tfrac{\sqrt{\tfrac{m\varpi
}{\pi\hbar}}}{2^{k}k!}\right)  ^{\frac{1}{2}}e^{-\tfrac{m\varpi}%
{2\hbar}\mu_1^2x_{1}^{2}}H_{k}^A\left(  \sqrt{\tfrac{m\varpi}{\hbar}}\mu_1x_{1}\right) \\
&\phi_{l}^B\left(  x_{2}\right)     =\left(  \tfrac{\sqrt{\tfrac{m\varpi}{\pi\hbar
}}}{2^{l}l!}\right)  ^{\frac{1}{2}}e^{-\tfrac{m\varpi}{2\hbar}%
\mu_2^2x_{2}^{2}}H_{l}^B\left(  \sqrt{\tfrac{m\varpi}{\hbar}}\mu_2x_{2}\right) \\
 &\chi_{m}^C\left(  x_{3}\right)    =\left(  \tfrac{\sqrt{\tfrac{m\varpi}{2\pi\hbar
}}}{2^{m}m!}\right)  ^{\frac{1}{2}}e^{-\tfrac{m\varpi}{2\hbar}%
\mu_2^3x_{3}^{2}}H_{m}^C\left(  \sqrt{\tfrac{m\varpi}{\hbar}}\mu_3x_{3}\right).
\end{align}
As usual, the orthogonality properties give rise to the  coefficients in question
\begin{equation}
A_{n_{1},n_{2},n_{3}}^{k,l,m}=\iiint\psi_{n_{1},n_{2},n_{3}}^{ABC}\left(
x_{1},x_{2},x_{3}\right) \ \varphi_{k}^A\left(  x_{1}\right)\  \phi_{l}^B\left(
x_{2}\right)\  \chi_{m}^C\left(  x_{3}\right)  dx_{1}dx_{2}dx_{3}%
\end{equation}
or equivalent after substitution
\begin{align}\label{AACC}
A_{n_{1},n_{2},n_{3}}^{k,l,m}=  &\frac{\left(
\frac{m\varpi}{\pi\hbar}\right)  ^{\frac{3}{2}} } 
{{\sqrt{2^{n_{1}+n_{2}+n_{3}+k+l+m}n_{1}!n_{2}!n_{3}! k!l!m!}}}
\iiint dx_{1}dx_{2}dx_{3}\
e^{-\tfrac{m\varpi}{2\hbar}\left(  q_{1}%
	^{2}+q_{2}^{2}+q_{3}^{2}+    \mu_{1}^{2}x_{1}^{2}+\mu_{2}^{2}x_{2}^{2}+\mu_{3}^{2}%
	x_{3}^{2} \right)  } \\
&   H_{n_{1}}^A\left(  \sqrt{\tfrac{m\varpi}{\hbar}}q_{1}\right)
H_{n_{2}}^B\left(  \sqrt{\tfrac{m\varpi}{\hbar}}q_{2}\right)  H_{n_{3}}^C\left(
\sqrt{\tfrac{m\varpi}{\hbar}}q_{3}\right)H_{k}^A\left(  \sqrt{\tfrac{m\varpi}{\hbar}}\mu_{1}x_{1}\right)
H_{l}^B\left(  \sqrt{\tfrac{m\varpi}{\hbar}}\mu_{2}x_{2}\right)  H_{m}^C\left(
\sqrt{\tfrac{m\varpi}{\hbar}}\mu_{3}x_{3}\right)   \nonumber.
\end{align}
To obtain $A_{n_{1},n_{2},n_{3}}^{k,l,m}$ explicitly, we use the Rodrigues formula for Hermite polynomials
\begin{equation}
H_{n}\left(  \omega x\right)  =\frac{d^{n}}{dt^{n}}\left.  e^{-t^{2}+2\omega
xt}\right\vert _{t=0}%
\end{equation}
and then \eqref{AACC} becomes
\begin{align}
A_{n_{1},n_{2},n_{3}}^{k,l,m}=  &  
\frac{\left(
	\frac{m\varpi}{\pi\hbar}\right)  ^{\frac{3}{2}} } 
{{\sqrt{2^{n_{1}+n_{2}+n_{3}+k+l+m}n_{1}!n_{2}!n_{3}! k!l!m!}}}
\iiint\tfrac{d^{n_{1}}}{dx^{n_{1}}}%
\tfrac{d^{n_{2}}}{dy^{n_{2}}}\tfrac{d^{n_{3}}}{dz^{n_{3}}}\tfrac{d^{k}}%
{dt^{k}}\tfrac{d^{l}}{ds^{l}}\tfrac{d^{m}}{dw^{m}}
\left\{  e^{-\tfrac{m\varpi}{\hbar
}\left(  \mu_{1}^{2}x_{1}^{2}+\mu_{2}^{2}x_{2}^{2}+\mu_{3}^{2}x_{3}%
^{2}\right)  }\right. \nonumber\\
&   e^{-x^{2}+2\sqrt{\tfrac{m\varpi}{\hbar}}\left[  \mu_{1}\cos
\theta\cos\phi x_{1}-\mu_{2}\left(  \sin\theta\sin\varphi+\cos\theta
\cos\varphi\sin\phi\right)  x_{2}-\mu_{3}\left(  \sin\theta\cos\varphi
-\cos\theta\sin\phi\sin\varphi\right)  x_{3}\right]  x}\nonumber\\
&   e^{-y^{2}+2\sqrt{\tfrac{m\varpi}{\hbar}}\left[ \mu_{1}\sin\phi
x_{1}+\mu_{2}\cos\phi\cos\varphi x_{2}-\mu_{3}\cos\phi\sin\varphi
x_{3}\right] y}\\
&   e^{-z^{2}+2\sqrt{\tfrac{m\varpi}{\hbar}}\left[  \mu_{1}\cos\phi
\sin\theta x_{1}+\mu_{2}\left(  \cos\theta\sin\varphi-\sin\theta\cos
\varphi\sin\phi\right)  x_{2}+\mu_{3}\left(  \cos\theta\cos\varphi+\sin
\theta\sin\phi\sin\varphi\right)  x_{3}\right]  z}\nonumber\\
&  \left.  \left.  e^{-t^{2}+2\sqrt{\tfrac{m\varpi}{\hbar}}\mu_{1}%
x_{1}t}e^{-s^{2}+2\sqrt{\tfrac{m\varpi}{\hbar}}\mu_{2}x_{2}s}e^{-w^{2}%
+2\sqrt{\tfrac{m\varpi}{\hbar}}\mu_{3}x_{3}w}\right\}  \right\vert
_{x,y,z,t,s,w=0}dx_{1}dx_{2}dx_{3}\nonumber.
\end{align}
After performing the integration over the variables $x_{1},x_{2}$ and $x_{3}$, we get
\begin{equation}
A_{n_{1},n_{2},n_{3}}^{k,l,m}=\tfrac{d^{n_{1}}}{dx^{n_{1}}}\tfrac{d^{n_{2}}%
}{dy^{n_{2}}}\tfrac{d^{n_{3}}}{dz^{n_{3}}}\tfrac{d^{k}}{dt^{k}}\tfrac{d^{l}%
}{ds^{l}}\tfrac{d^{m}}{dw^{m}}\left.  \tfrac{e^{2a_{1}tx}e^{2b_{1}ty}%
e^{2c_{1}tz}e^{2a_{2}sx}e^{2b_{2}sy}e^{2c_{2}sz}e^{2a_{3}wx}e^{2b_{3}%
wy}e^{2c_{3}wz}}{\sqrt{2^{n_{1}+n_{2}+n_{3}+k+l+m}n_{1}!n_{2}!n_{3}!k!l!m!}%
}\right\vert _{x,y,z,t,s,w=0}\label{32}%
\end{equation}
where we have set the following parameters
\begin{align}	
&a_{1}  =\cos\theta\cos\phi\label{333}\\
&b_{1}   =\sin\phi \label{334}
\\
&c_{1}   =\cos\phi\sin\theta \label{335}
\\ &
a_{2}  =-\sin\theta\sin\varphi-\cos\theta\cos\varphi\sin\phi\label{336}\\
&b_{2}  =\cos\phi\cos\varphi \label{337}
\\
& c_{2}  =\cos\theta\sin\varphi-\sin\theta\cos\varphi\sin\phi \label{338}
\\ &
a_{3}  =\cos\theta\sin\phi\sin\varphi-\sin\theta\cos\varphi\label{339}\\
& b_{3}  =-\cos\phi\sin\varphi \label{330}
\\
& c_{3}   =\cos\theta\cos\varphi+\sin\theta\sin\phi\sin\varphi
\label{331}.
\end{align}
As we will see, the above set will play a crucial role in determining the purify functions for different configurations, depending on how the interaction between oscillators is treated. We can write \eqref{32} in a different form by expanding each exponential as a power series to end up with the following expression
\begin{align}
A_{n_{1},n_{2},n_{3}}^{k,l,m} = & \tfrac{d^{n_{1}}}{dx^{n_{1}}}\tfrac
{d^{n_{2}}}{dy^{n_{2}}}\tfrac{d^{n_{3}}}{dz^{n_{3}}}\tfrac{d^{k}}{dt^{k}%
}\tfrac{d^{l}}{ds^{l}}\tfrac{d^{m}}{dw^{m}}
{\sum_{i_{1}, \cdots,i_{9}=0}^{\infty}}
\Big\{  \tfrac{t^{i_{1}+i_{2}+i_{3}}s^{i_{4}+i_{5}+i_{6}}w^{i_{7}+i_{8}%
+i_{9}}x^{i_{1}+i_{4}+i_{7}}y^{i_{2}+i_{5}+i_{8}}z^{i_{3}+i_{6}+i_{9}}}%
{\sqrt{2^{n_{1}+n_{2}+n_{3}+k+l+m}n_{1}!n_{2}!n_{3}!k!l!m!}}
  \nonumber\\ &   \left. 
\tfrac{\left(
2a_{1}\right)  ^{i_{1}}}{i_{1}!}\tfrac{\left(  2b_{1}\right)  ^{i_{2}}}%
{i_{2}!}
\tfrac{\left(  2c_{1}\right)  ^{i_{3}}}{i_{3}!}%
\tfrac{\left(  2a_{2}\right)  ^{i_{4}}}{i_{4}!}\tfrac{\left(  2\allowbreak
b_{2}\right)  ^{i_{5}}}{i_{5}!}\tfrac{\left(  2c_{2}\right)  ^{i_{6}}}{i_{6}%
!}\tfrac{\left(  2a_{3}\right)  ^{i_{7}}}{i_{7}!}\tfrac{\left(  2\allowbreak
b_{3}\right)  ^{i_{8}}}{i_{8}!}\tfrac{\left(  2c_{3}\right)  ^{i_{9}}}{i_{9}%
!}\Big\}  \right\vert _{x,y,z,t,s,w=0}\label{aa}.
\end{align}
By using the identity $\left. \frac{d^{k}}{dt^{k}}t^{i}\right\vert_{t=0}=k!\delta_{k,i}$ and taking derivatives with respect to the variables ($t, s, w, x, y, z$), we demonstrate the result
\begin{align}\label{AASK}
A_{n_{1},n_{2},n_{3}}^{k,l,m}   =&
{\sum_{i_{1}, \cdots,i_{9}=0}^{\infty}}
  \tfrac{\sqrt{n_{1}!n_{2}!n_{3}!k!l!m!}\ \left(  a_{3}\right)  ^{n_{1}%
}\left(  \allowbreak b_{3}\right)  ^{n_{2}}\left(  c_{1}\right)  ^{k}\left(
c_{2}\right)  ^{l}\left(  c_{3}\right)  ^{n_{3}-k-l}}{\left(  n_{3}%
-k-l+i_{1}+i_{2}+i_{3}+i_{4}\right)  !} \tfrac{\left(  \frac{a_{2}c_{3}}{c_{2}a_{3}}\right)  ^{i_{3}%
}\left(  \frac{a_{1}c_{3}}{c_{1}a_{3}}\right)  ^{i_{1}}\left(  \frac
{b_{1}c_{3}}{c_{1}b_{3}}\right)  ^{i_{2}}\left(  \allowbreak\frac{b_{2}c_{3}%
}{c_{2}b_{3}}\right)  ^{i_{4}}}{\left(  n_{2}-i_{2}-i_{4}\right)  !\left(
l-i_{3}-i_{4}\right)  !\left(  k-i_{1}-i_{2}\right)  !\left(  n_{1}%
-i_{1}-i_{3}\right)  !i_{1}!i_{2}!i_{3}!i_{4}!} \nonumber\\
& \delta_{i_{1}+i_{2}+i_{3},k}\ \delta_{i_{4}+i_{5}+i_{6},l}\ \delta
_{i_{7}+i_{8}+i_{9},m}\ \delta_{i_{1}+i_{4}+i_{7},n_{1}}\ \delta_{i_{2}%
+i_{5}+i_{8},n_{2}}\ \delta_{i_{3}+i_{6}+i_{9},n_{3}}.  
\end{align}
At this level, we introduce the Exton's $K_{16}$ hypergeometric function \cite{Exton} 
\begin{equation}
K_{16}\left(  \alpha_{1},\alpha_{2},\alpha_{3},\alpha_{4};\beta
;x,y,z,t\right)  =
{\textstyle\sum\limits_{m_{1},m_{2},m_{3},m_{4}=0}^{\infty}}
\tfrac{\left(  \alpha_{1}\right)  _{m_{1}+m_{2}}\left(  \alpha_{2}\right)
_{m_{2}+m_{3}}\left(  \alpha_{3}\right)  _{m_{3}+m_{4}}\left(  \alpha
_{4}\right)  _{m_{4}+m_{1}}}{\left(  \beta\right)  _{m_{1}+m_{2}+m_{3}+m_{4}}%
}\tfrac{x^{m_{1}}y^{m_{2}}z^{m_{3}}t^{m_{4}}}{m_{1}!m_{2}!m_{3}!m_{4}!}%
\end{equation}
where $\left(  a\right)  _{k}$ is the Pochhammer symbol 
$\left(  a\right)  _{k}=\tfrac{\Gamma\left(  a+k\right)  }{\Gamma\left(
a\right)  }=\left(  -1\right)  ^{k}\tfrac{\Gamma\left(  1-a\right)  }%
{\Gamma\left(  1-a-k\right)  }$. Consequently, we can easily show that \eqref{AASK} can be expressed in terms of $K_{16}$ as
\begin{align}
A_{n_{1},n_{2},n_{3}}^{k,l,m}= & \ \delta_{m,n_{1}+n_{2}+n_{3}-k-l}   
\tfrac{\sqrt{n_{1}!n_{2}!n_{3}!k!l!\left(  n_{1}+n_{2}+n_{3}-k-l\right)
!}\left(  a_{3}\right)  ^{n_{1}}\left(  \allowbreak b_{3}\right)  ^{n_{2}%
}\left(  c_{1}\right)  ^{k}\left(  c_{2}\right)  ^{l}\left(  c_{3}\right)
^{n_{3}-k-l}}{\left(  n_{3}-k-l\right)  !n_{1}!k!n_{2}!l!}  \nonumber\\
&    K_{16}\left(  -n_{1},-k,-n_{2},-l;n_{3}-k-l+1;\tfrac
{a_{2}c_{3}}{c_{2}a_{3}},\tfrac{a_{1}c_{3}}{c_{1}a_{3}},\tfrac{b_{1}c_{3}%
}{c_{1}b_{3}},\tfrac{b_{2}c_{3}}{c_{2}b_{3}}\right)  
\end{align}
and then, to have non-zero coefficients, the condition $m+k+l=n_{1}+n_{2}+n_{3}$ should be satisfied. Note that $K_{16}$ now becomes a 4-variable polynomial of order $n_1,k,n_2,l$.  As a result we get
\begin{align}\label{Afin}
A_{n_{1},n_{2},n_{3}}^{k,l}= &  \sum\limits_{m}A_{n_{1},n_{2},n_{3}}%
^{k,l,m}\delta_{m,n_{1}+n_{2}+n_{3}-k-l}\\
=&  \tfrac{\sqrt{n_{1}%
!n_{2}!n_{3}!k!l!\left(  n_{1}+n_{2}+n_{3}-k-l\right)  !}\left(  a_{3}\right)
^{n_{1}}\left(  \allowbreak b_{3}\right)  ^{n_{2}}\left(  c_{1}\right)
^{k}\left(  c_{2}\right)  ^{l}\left(  c_{3}\right)  ^{n_{3}-k-l}}{\left(
n_{3}-k-l\right)  !n_{1}!k!n_{2}!l!}  \nonumber\\
&   K_{16}\left(  -n_{1},-k,-n_{2},-l;n_{3}-k-l+1;\tfrac
{a_{2}c_{3}}{c_{2}a_{3}},\tfrac{a_{1}c_{3}}{c_{1}a_{3}},\tfrac{b_{1}c_{3}%
}{c_{1}b_{3}},\tfrac{b_{2}c_{3}}{c_{2}b_{3}}\right)  
\end{align}
Plugging this into \eqref{244} to write the wavefunctions as
\begin{equation}
\psi_{n_{1},n_{2},n_{3}}^{ABC}\left(  x_{1},x_{2},x_{3}\right)  =\sum\limits_{k,l}A_{n_{1},n_{2},n_{3}}^{k,l}\ \varphi_{k}^A\left(
x_{1}\right)\  \phi_{l}^B\left(  x_{2}\right)  \chi_{n_{1}+n_{2}+n_{3}%
-k-l}^C\left(  x_{3}\right)  \label{533}.
\end{equation}
Having derived \eqref{533} together with \eqref{AASK}, we will now explore its application in the following steps. In particular, we will use the associated density matrices to determine the purity functions and to study the fundamental properties of entanglement.

\section{Entanglement analysis}\label{Entag} 

We will begin our analysis of entanglement in the present system by starting with the general case and then examining more specific cases. Considering both the general situation and specific examples will help to demonstrate our results and highlight various fundamental properties that depend on the values of the physical parameters involved. Considering both broad and narrow scenarios will help to illustrate how entanglement is affected by different conditions in the system and allow us to highlight the basic features that emerge under different parametric conditions. This approach, moving from the general to the specific, will provide insight into how the nature of entanglement changes with the dynamical settings governing the system.

\subsection{Generic case}

There are three different ways to establish the Schmidt decomposition of the equation \eqref{533}, based on the following interaction configurations \cite{Sudbery}: 

1) $A$-$BC$ interaction: 
The subsystem $A$ is considered separately from the common subsystem $BC$. 

2) $B$-$AC$ interaction:  
The subsystem $B$ is considered separately from the common subsystem $AC$.

3) $C$-$AB$ interaction:
The subsystem $C$ is considered separately from the common subsystem $AB$.

These three ways of applying the Schmidt decomposition, corresponding to the different interaction arrangements, are denoted as
\begin{align}
\psi_{n_{1},n_{2},n_{3}}^{ABC}\left(  x_{1},x_{2},x_{3}\right)   &  =
{\textstyle\sum\limits_{k=0}^{n_{1}+n_{2}+n_{3}}}
\sqrt{\alpha_{k}}\ \varphi_{k}^{A}\left(  x_{1}\right)  \ \Theta_{k}^{BC}\left(
x_{2},x_{3}\right)  \label{544}\\
&  =%
{\textstyle\sum\limits_{l=0}^{n_{1}+n_{2}+n_{3}}}
\sqrt{\beta_{l}}\ \phi_{l}^{B}\left(  x_{2}\right)\  \Phi_{l}^{AC}\left(
x_{1},x_{3}\right) \label{545} \\
&  =
{\textstyle\sum\limits_{\sigma=0}^{n_{1}+n_{2}+n_{3}}}
\sqrt{\gamma_{\sigma}}\ \chi_{\sigma}^{C}\left(  x_{3}\right)\  \Xi_{\sigma}%
^{AB}\left(  x_{1},x_{2}\right)\label{546}
\end{align}
where the sets $\left\{ \phi_{k}^A( x_{1})\right\}, \left\{\theta
_{l}^B( x_{2}) \right\}$ and $\left\{\chi_{\sigma}^C( x_{3}) \right\}$ consist of orthonormal pairs of single particle states, while the sets $\left\{\Theta_{k}^{BC}\left(
x_{2},x_{3}\right) \right\}, \left\{\Phi_{l}^{AC}\left(
x_{1},x_{3}\right) \right\}$ and $\left\{\Xi_{\sigma}
^{AB}\left( x_{1},x_{2}\right) \right\}$ contain orthonormal pairs of two-particle states defined by
\begin{align}
\Theta_{k}^{BC}\left(  x_{2},x_{3}\right)    & =%
{\textstyle\sum\limits_{l=0}^{n_{1}+n_{2}+n_{3}-k}}
\phi_{l}\left(  x_{2}\right) \ \chi_{n_{1}+n_{2}+n_{3}-k-l}\left(
x_{3}\right)  \label{A-BC}\\
\Phi_{l}^{AC}\left(  x_{1},x_{3}\right)    & =%
{\textstyle\sum\limits_{k=0}^{n_{1}+n_{2}+n_{3}-l}}
\varphi_{k}\left(  x_{1}\right) \ \chi_{n_{1}+n_{2}+n_{3}-k-l}\left(
x_{3}\right)  \label{B-AC}\\
\Xi_{\sigma}^{AB}\left(  x_{1},x_{2}\right)    & =%
{\textstyle\sum\limits_{k=0}^{n_{1}+n_{2}+n_{3}-\sigma}}
\varphi_{k}\left(  x_{1}\right) \ \phi_{n_{1}+n_{2}+n_{3}-k-\sigma}\left(
x_{2}\right)  \label{C-AB}.
\end{align}
Here the parameters $\alpha_{k}\equiv\alpha_{k}^{n_{1},n_{2},n_{3}}$, $\beta_{l}\equiv\beta_{l}^{n_{1},n_{2},n_{3}}$ and $\gamma_{\sigma}\equiv\gamma_{\sigma}^{n_{1},n_{2},n_{3}}$ are functions of the Schmidt modes, as demonstrated in the context below
\begin{align}
\label{alpha} \alpha_{k} &  =
{\sum_{l=0}^{n_{1}+n_{2}+n_{3}-k}}
\left(  A_{n_{1},n_{2},n_{3}}^{k,l}\right)  ^{2}\\
\label{beta} \beta_{l} &  =
{\sum_{k=0}^{n_{1}+n_{2}+n_{3}-l}}
\left(  A_{n_{1},n_{2},n_{3}}^{k,l}\right)  ^{2}\\
\label{gamma} \gamma_{\sigma} &  =
{\sum_{k=0}^{n_{1}+n_{2}+n_{3}-\sigma}}
\left(  A_{n_{1},n_{2},n_{3}}^{k,n_{1}+n_{2}+n_{3}-k-\sigma}\right)  ^{2}
\end{align}
which are nothing but the eigenvalues of the reduced density matrices $\left(\rho^{A},\rho^{BC}\right)$, $\left(\rho^{B},\rho^{AC}\right)$, and $\left(\rho^{C},\rho^{AB}\right)$, respectively, given by
\begin{align}
\rho^{A}  &  =
{\textstyle\sum\limits_{k=0}^{n_{1}+n_{2}+n_{3}}}
\alpha_{k}\ \varphi_{k}^{A}\left(  x_{1}\right) \ \varphi_{k}^{\ast A}\left(
x_{1}^{\prime}\right) \\
\rho^{BC}  &  =
{\textstyle\sum\limits_{k=0}^{n_{1}+n_{2}+n_{3}}}
\alpha_{k}\ \Theta_{k}^{BC}\left(  x_{2},x_{3}\right) \ \Theta_{k}^{\ast
	BC}\left(  x_{2}^{\prime},x_{3}^{\prime}\right)\\
\rho^{B}  &  =
{\textstyle\sum\limits_{l=0}^{n_{1}+n_{2}+n_{3}}}
\beta_{l}\ \phi_{l}^{B}\left(  x_{2}\right) \  \phi_{l}^{\ast B}\left(
x_{2}^{\prime}\right) \\
\rho^{AC}  &  =
{\textstyle\sum\limits_{l=0}^{n_{1}+n_{2}+n_{3}}}
\beta_{l}\ \Phi_{l}^{AC}\left(  x_{1},x_{3}\right)\  \Phi_{l}^{\ast AC}\left(
x_{1}^{\prime},x_{3}^{\prime}\right) \\
\rho^{C}  &  =
{\textstyle\sum\limits_{\sigma=0}^{n_{1}+n_{2}+n_{3}}}
\gamma_{\sigma}\ \chi_{\sigma}^{C}\left(  x_{3}\right) \ \chi_{\sigma}^{\ast
	C}\left(  x_{3}\right)\\
\rho^{AB}  &  =
{\textstyle\sum\limits_{\sigma=0}^{n_{1}+n_{2}+n_{3}}}
\gamma_{\sigma}\ \Xi_{\sigma}^{AB}\left(  x_{1},x_{2}\right) \ \Xi_{\sigma}^{\ast
	AB}\left(  x_{1}^{\prime},x_{2}^{\prime}\right).
\end{align}
From these densities we can derive the corresponding purity functions ${ P}_{n_{1},n_{2},n_{3}}^{A}\left(\theta,\varphi,\phi\right) $, ${ P}_{n_{1},n_{2},n_{3}}^{B}\left(\theta,\varphi,\phi\right) $ and ${ P}_{n_{1},n_{2},n_{3}}^{C}\left(\theta,\varphi,\phi\right) $. Indeed, this will be the focus of our upcoming study, where we will explore intriguing configurations of the associated quantum numbers \((n_1, n_2, n_3)\), with a particular emphasis on entanglement as a core aspect of the investigation.

\subsection{A-BC interaction}

We consider the first configuration, which is the {\bf A-BC} interaction described by 
 the following Schmidt decomposition  as given in \eqref{544}
\begin{align}
	\psi_{n_{1},n_{2},n_{3}}^{ABC}\left(x_{1},x_{2},x_{3}\right)    ={\textstyle\sum\limits_{k=0}^{n_{1}+n_{2}+n_{3}}}\sqrt{\alpha_{k}}\ \varphi_{k}^{A}\left(x_{1}\right)\ \Theta_{k}^{BC}\left(x_{2},x_{3}\right)  \label{5444}    
	\end{align}
and analyze three different scenarios: \((n_{1}=0, n_{2}=0, n_{3}\neq 0)\), \((n_{1}=0, n_{2}\neq 0, n_{3}=0)\), and \(( n_{1}\neq 0, n_{2}=0, n_{3}=0)\), with the goal of identifying their primary differences.

	\subsubsection*{\bf B1. \ A-BC case 1: $n_{1}=0, n_{2}=0, n_{3}\neq 0$}
	
	 After replacing $n_{1}=0,n_{2}=0,$ in \eqref{alpha}, and using \eqref{Afin} together with \eqref{335}, \eqref{338}, \eqref{331}, we get the Schmidt modes
\begin{align}
	\alpha_{k}^{0,0,n_{3}} =
	\tfrac{n_{3}!}{k!\left(
		n_{3}-k\right)  !} \sin^{2k}\theta \cos^{2k} \phi\
		\left(1-\sin^{2}\theta \cos^{2} \phi\right)^{n_3-k}
\end{align}
which allows us to easily verify the von Neumann normalization condition
\begin{equation}
	\tr\rho_{0,0,n_{3}}^{A}=
	\tr\rho_{0,0,n_{3}}^{BC}=%
	{\sum_{k=0}^{n_{3}}}
	\alpha_{k}^{0,0,n_{3}}
	=1.
\end{equation}
In general,  the purity function is defined as the trace of the squared
density operator. Since  ${P}_{n_{1},n_{2},n_{3}}^{A}\left(  \theta,\varphi,\phi\right)  ={P}_{n_{1}%
	,n_{2},n_{3}}^{BC}\left(  \theta,\varphi,\phi\right) $, then we write
\begin{equation}\label{Pure}
	{P}_{n_{1},n_{2},n_{3}}^{A}\left(  \theta,\varphi,\phi\right)  =
		\tr\left(  \rho
	_{n_{1},n_{2},n_{3}}^{A}\right)  ^{2}=
	{\sum\limits_{k=0}^{n_{1}+n_{2}+n_{3}}}
	\left(  \alpha_{k}^{n_{1},n_{2},n_{3}}\right)  ^{2}%
\end{equation}
and for the configuration $ n_{1}=n_{2}=0, n_{3} $, we get
\begin{align}\label{PA00n3}
{P}_{0,0,n_{3}}^{A}\left(  \theta,\varphi,\phi\right)    & =
	{\sum_{k=0}^{n_{3}}}
	\left(  \tfrac{n_{3}!}{k!\left(  n_{3}-k\right)  !}\right)  ^{2}
	c_{1}^{4k}\left(   c_{3}^{2}+c_{2}^{2}\right)
	^{2(n_{3}-k)}   
\\
	& =\left(  \left(  \cos^{2}\theta+\sin^{2}\theta\sin^{2}\phi\right)  ^{2}%
	-\sin^{4}\theta\cos^{4}\phi\right)  ^{n_{3}}{\mathcal P}_{n_{3}}\left(  \frac{\left(
		\cos^{2}\theta+\sin^{2}\theta\sin^{2}\phi\right)  ^{2}+\sin^{4}\theta\cos
		^{4}\phi}{\left(  \cos^{2}\theta+\sin^{2}\theta\sin^{2}\phi\right)  ^{2}%
		-\sin^{4}\theta\cos^{4}\phi}\right)
\end{align}
and we have used the following identity  \cite{Prudnikov}
\begin{align}
{\sum_{k=0}^{n}}
\left(  \tfrac{n!}{k!\left(  n-k\right)  !}\right)  ^{2}x^{k}=\left(
1-x\right)  ^{n}{\mathcal P}_{n}\left(  \frac{1+x}{1-x}\right)
\end{align}
where ${\mathcal P}_{n}$ are the Legendre polynomials of order $n$ and  the four first ones are given by
\begin{equation}
	{\mathcal P}_{0}\left(  z\right)  =1,\quad {\mathcal P}_{1}\left(  z\right)  =z,\quad {\mathcal P}_{2}\left(  z\right)
	=\frac{1}{2}\left(  3z^{2}-1\right), \quad {\mathcal P}_{3}\left(  z\right)  =\frac{1}
	{2}\left(  5z^{3}-3z\right).
\end{equation}

\subsubsection*{\bf B2. \ A-BC case 2: $n_{1}=0, n_{2}\neq0, n_{3}= 0$}

As before, we consider \eqref{alpha} with the condition $n_{1}=0,n_{3}=0,$ and \eqref{Afin} as well as \eqref{334}, \eqref{337}, \eqref{330} to get the Schmidt modes
\begin{align}
	\alpha_{k}^{0,n_{2},0}=
	{\sum_{l=0}^{n_{2}-k}}
	\left(  A_{0,n_{2},0}^{k,l}\right)  ^{2}
		=\tfrac{n_{2}!\left(  b_{1}^{2}\right)  ^{k}}{k!\left(
		n_{2}-k\right)  !}\left(  b_{3}^{2}+b_{2}^{2}\right)  ^{n_{2}-k}%
\end{align}
showing the von Neumann normalization condition
\begin{equation}
	\tr\rho_{0,n_{2},0}^{A}=\tr\rho_{0,n_{2},0}^{BC}=%
	{\sum_{k=0}^{n_{2}}}
	\alpha_{k}^{0,n_{2},0}=\left(  b_{1}^{2}+b_{3}^{2}+b_{2}^{2}\right)  ^{n_{2}%
	}=1.
\end{equation}
From \eqref{Pure}
with $\ n_{1}=0, n_{3}=0  $, we
obtain
\begin{align}
	P_{0,n_{2},0}^{A}\left(  \theta,\varphi,\phi\right)   
	& =\left(  \left(  b_{3}^{2}+b_{2}^{2}\right)  ^{2}-b_{1}^{4}\right)  ^{n_{2}
	}{\mathcal P}_{n_{2}}\left(  \frac{\left(  b_{3}^{2}+b_{2}^{2}\right)  ^{2}+b_{1}^{4}
	}{\left(  b_{3}^{2}+b_{2}^{2}\right)  ^{2}-b_{1}^{4}}\right)  \\
	& =  \cos^{n_{2}}(2\phi) {\mathcal P}_{n_{2}}\left(  \frac{
		\cos\left(  4\phi\right)  +3}{4\cos2\phi}\right)
\end{align}
For $n_2=1,2$, we get
\begin{align}
\label{P010} &P_{0,1,0}^{A}\left(  \theta,\varphi,\phi\right)  =\cos2\phi
\\
& \label{P020}
	P_{0,2,0}^{A}\left(  \theta,\varphi,\phi\right)    
	=\frac{5}{16}\cos4\phi+\frac{3}{64}\cos8\phi+\frac{41}{64}%
\end{align} 
By comparing these purities with those derived in \eqref{P001} and \eqref{P002}, we can identify the factors that contribute to the observed differences. Consequently, we conclude that the quantum numbers are crucial in producing the different results.

	\subsubsection*{\bf B3. \ A-BC case  3: $n_{1}\neq0, n_{2}=0, n_{3}= 0$}

To complete our study, we analyze the last case and highlight the factors that distinguish it from the two previous ones. Then, from 
\eqref{alpha} with $n_{2}=0,n_{3}=0,$, and \eqref{Afin} as well as \eqref{333}, \eqref{336}, \eqref{339}, we find the Schmidt modes
\begin{equation}
	\alpha_{k}^{n_{1},0,0}=
	{\sum_{l=0}^{n_{1}-k}}
	\left(  A_{n_{1},0,0}^{k,l}\right)  ^{2}
		=
		\tfrac{n_{1}!\left(  a_{1}^{2}\right)  ^{k}}{k!\left(
		n_{1}-k\right)  !}\left(  a_{3}^{2}+a_{2}^{2}\right)  ^{n_{1}-k}%
\end{equation}
which leads to the von Neumann normalization condition
\begin{equation}
	\tr\rho_{n_{1},0,0}^{A}=\tr\rho_{n_{1},0,0}^{BC}=%
	{\sum_{k=0}^{n_{1}}}
	\alpha_{k}^{n_{1},0,0}=\left(  a_{1}^{2}+a_{3}^{2}+a_{2}^{2}\right)  ^{n_{1}%
	}=1.
\end{equation}
By replacing $n_{2}=0,n_{3}=0$ into \eqref{Pure}, we get the following purity
\begin{align}
	P_{n_{1},0,0}^{A}\left(  \theta,\varphi,\phi\right)    & =
	\left(  \left(  a_{3}^{2}+a_{2}^{2}\right)  ^{2}-a_{1}^{4}\right)
	^{n_{1}}{\mathcal P}_{n_{1}}\left(  \frac{\left(  a_{3}^{2}+a_{2}^{2}\right)  ^{2}%
		+a_{1}^{4}}{\left(  a_{3}^{2}+a_{2}^{2}\right)  ^{2}-a_{1}^{4}}\right)  \\
	& =\left(  \left(  \sin^{2}\theta+\cos^{2}\theta\sin^{2}\phi\right)  ^{2}%
	-\cos^{4}\theta\cos^{4}\phi\right)  ^{n_{3}}{\mathcal P}_{n_{1}}\left(  \frac{\left(
		\sin^{2}\theta+\cos^{2}\theta\sin^{2}\phi\right)  ^{2}+\cos^{4}\theta\cos
		^{4}\phi}{\left(  \sin^{2}\theta+\cos^{2}\theta\sin^{2}\phi\right)  ^{2}%
		-\cos^{4}\theta\cos^{4}\phi}\right).
\end{align}
In particular, for $n_1=1,2$, we have
\begin{align}
	&P_{1,0,0}^{A}\left(  \theta,\varphi,\phi\right)  =\left(  \sin^{2}\theta
	+\cos^{2}\theta\sin^{2}\phi\right)  ^{2}+\cos^{4}\theta\cos^{4}\phi
	\\
	&P_{2,0,0}^{A}\left(  \theta,\varphi,\phi\right)  =\frac{1}{2}\left(  3\left(
	\left(  \sin^{2}\theta+\cos^{2}\theta\sin^{2}\phi\right)  ^{2}+\cos^{4}%
	\theta\cos^{4}\phi\right)  ^{2}-\left(  \left(  \sin^{2}\theta+\cos^{2}%
	\theta\sin^{2}\phi\right)  ^{2}-\cos^{4}\theta\cos^{4}\phi\right)
	^{2}\right)
\end{align}

\subsection{B-AC interaction}

As for the second configuration, we consider the {\bf B-AC} interaction to have the  Schmidt decomposition resulting from \eqref{545}
\begin{align}
	\psi_{n_{1},n_{2},n_{3}}^{ABC}\left(  x_{1},x_{2},x_{3}\right)    
	=   {\textstyle\sum\limits_{l=0}^{n_{1}+n_{2}+n_{3}}}
	\sqrt{\beta_{l}}\ \phi_{l}^{B}\left(  x_{2}\right)\  \Phi_{l}^{AC}\left(
	x_{1},x_{3}\right) \label{5455}  
\end{align}
and we proceed to analyze three different scenarios: \((n_{1}=0, n_{2}=0, n_{3}\neq 0)\), \((n_{1}=0, n_{2}\neq 0, n_{3}=0)\), and \(( n_{1}\neq 0, n_{2}=0, n_{3}=0)\).

\subsubsection*{\bf C1. \ B-AC case 1: $n_{1}=0, n_{2}=0, n_{3}\neq 0$}

Substituting \(n_{1}=0, n_{2}=0\) into \eqref{beta} and using \eqref{Afin} together with \eqref{335}, \eqref{338}, \eqref{331}, we get the Schmidt modes
\begin{align}
\beta_{l}^{0,0,n_{3}}&=
{\sum_{k=0}^{n_{3}-l}}
\left(  A_{0,0,n_{3}}^{k,l}\right)  ^{2}
	\\
&	=\tfrac{n_{3}!\left(  c_{2}^{2}\right)  ^{l}}{l!\left(
	n_{3}-l\right)  !}\left(  c_{3}^{2}+c_{1}^{2}\right)  ^{n_{3}-l}%
\end{align}
with the condition
\begin{equation}
	\tr\rho_{0,0,n_{3}}^{B}=\tr\rho_{0,0,n_{3}}^{AC}=
	{\sum_{l=0}^{n_{3}}}
	\beta_{l}^{0,0,n_{3}}=\left(  c_{2}^{2}+c_{1}^{2}+c_{3}^{2}\right)  ^{n_{3}}=1.
\end{equation}
We show that the corresponding purity is given by
\begin{align}\label{PB00n3}
	P_{0,0,n_{3}}^{B}\left(  \theta,\varphi,\phi\right)     
	=&
	\left(  \tfrac{n_{3}!}{l!\left(  n_{3}-l\right)  !}\right)  ^{2}\left(
	c_{2}^{2}\right)  ^{l}\left(  c_{3}^{2}+c_{1}^{2}\right)  ^{n_{3}-l}=\left(
	\left(  c_{1}^{2}+c_{3}^{2}\right)  ^{2}-c_{2}^{4}\right)  ^{n_{3}}{\mathcal P}_{n_{3}%
	}\left(  \frac{\left(  c_{1}^{2}+c_{3}^{2}\right)  ^{2}+c_{2}^{4}}{\left(
		c_{1}^{2}+c_{3}^{2}\right)  ^{2}-c_{2}^{4}}\right)  \\
	 =&\left(  \cos^{2}\phi\sin^{2}\theta+\left(  \cos\theta\cos\varphi+\sin
	\theta\sin\phi\sin\varphi\right)  ^{2}-\cos^{2}\theta\sin^{2}\varphi\right)
	^{n_{3}}\\
	&  {\mathcal P}_{n_{3}}\left(  \frac{\left(  \cos^{2}\phi\sin^{2}\theta+\left(
		\cos\theta\cos\varphi+\sin\theta\sin\phi\sin\varphi\right)  ^{2}\right)
		^{2}+\left(  \cos\theta\sin\varphi-\sin\theta\cos\varphi\sin\phi\right)  ^{4}%
	}{\cos^{2}\phi\sin^{2}\theta+\left(  \cos\theta\cos\varphi+\sin\theta\sin
		\phi\sin\varphi\right)  ^{2}-\cos^{2}\theta\sin^{2}\varphi}\right)
		\nonumber.
\end{align}

\subsubsection*{\bf C2. \ B-AC case 2: $n_{1}=0, n_{2}\neq0, n_{3}= 0$}

As mentioned before, we analyze \eqref{beta} under the conditions \( n_{1} = 0 \) and \( n_{3} = 0 \), along with \eqref{Afin} and equations \eqref{333}, \eqref{337}, and \eqref{330} to obtain the Schmidt modes
	\begin{align}
		\beta_{k}^{0,n_{2},0} &=
		{\sum_{l=0}^{n_{2}-k}}
		\left(  A_{0,n_{2},0}^{k,l}\right)  ^{2}\\
		&=\tfrac{n_{2}!\left(  b_{2}%
			^{2}\right)  ^{k}}{k!\left(  n_{2}-k\right)  !}\left(  b_{1}^{2}+b_{3}%
		^{2}\right)  ^{n_{2}-k}%
	\end{align}
	leading to the condition%
	\begin{equation}
		\tr\rho_{0,n_{2},0}^{B}=\tr\rho_{0,n_{2},0}^{AC}=
		{\sum_{k=0}^{n_{2}}}
		\beta_{k}^{0,n_{2},0}=\left(  b_{1}^{2}+b_{3}^{2}+b_{2}^{2}\right)  ^{n_{2}}=1
	\end{equation}
The associated purity reads as
	\begin{align}
		P_{0,n_{2},0}^{B}\left(  \theta,\varphi,\phi\right)   
		 & =
		\left(  \left(  b_{1}^{2}+b_{3}%
		^{2}\right)  ^{2}-b_{2}^{4}\right)  ^{n_{2}}{\mathcal P}_{n_{2}}\left(  \frac{\left(
			b_{1}^{2}+b_{3}^{2}\right)  ^{2}+b_{2}^{4}}{\left(  b_{1}^{2}+b_{3}%
			^{2}\right)  ^{2}-b_{2}^{4}}\right)  \\
		& =\left(  \sin^{2}\phi-\cos^{2}\varphi\cos^{2}\phi+\cos^{2}\phi\sin
		^{2}\varphi\right)  ^{n_{2}}{\mathcal P}_{n_{2}}\left(  \frac{\left(  \sin^{2}\phi
			+\cos^{2}\phi\sin^{2}\varphi\right)  ^{2}+\cos^{4}\phi\cos^{4}\varphi}%
		{\sin^{2}\phi-\cos^{2}\varphi\cos^{2}\phi+\cos^{2}\phi\sin^{2}\varphi}\right)
	\end{align}
	and for $n_2=1,2$, we get
	\begin{align}
	P_{0,1,0}^{B}\left(  \theta,\varphi,\phi\right)  =&\left(  \sin^{2}\phi
		+\cos^{2}\phi\sin^{2}\varphi\right)  ^{2}+\cos^{4}\phi\cos^{4}\varphi
	\\
		P_{0,2,0}^{A}\left(  \theta,\varphi,\phi\right)     =&\frac{3}{2}\left(
		\left(  \sin^{2}\phi+\cos^{2}\phi\sin^{2}\varphi\right)  ^{2}+\cos^{4}\phi
		\cos^{4}\varphi\right)  ^{2}-\frac{1}{2}\left(  \sin^{2}\phi-\cos^{2}\varphi\cos^{2}\phi+\cos^{2}%
		\phi\sin^{2}\varphi\right)  ^{2}%
	\end{align}%

\subsubsection*{\bf C3. \ B-AC case 3: $n_{1}\neq 0, n_{2}=0, n_{3}= 0$}

From  \eqref{beta} with  $n_{2}=0,n_{3}=0,$ by using \eqref{Afin} and \eqref{333}, \eqref{336}, \eqref{339}, we end up with the Schmidt modes
\begin{align}
	\beta_{k}^{n_{1},0,0} &=
	{\sum_{l=0}^{n_{1}-k}}
	\left(  A_{n_{1},0,0}^{k,l}\right)  ^{2}\\
		&=\tfrac{n_{1}!\left(  a_{2}^{2}\right)  ^{k}}{k!\left(
		n_{1}-k\right)  !}\left(  a_{1}^{2}+a_{3}^{2}\right)  ^{n_{1}-k}%
\end{align}
and condition is met
\begin{equation}
	\tr\rho_{n_{1},0,0}^{B}=\tr\rho_{n_{1},0,0}^{AC}=%
	{\sum_{k=0}^{n_{1}}}
	\beta_{k}^{n_{1},0,0}=\left(  a_{1}^{2}+a_{3}^{2}+a_{2}^{2}\right)  ^{n_{1}}=1.
\end{equation}
As a result, we obtain purity
\begin{align}
	P_{n_{1},0,0}^{B}\left(  \theta,\varphi,\phi\right)     =&
	\left(  \left(  a_{1}^{2}+a_{3}^{2}\right)  ^{2}-a_{2}^{4}\right)
	^{n_{1}}{\mathcal P}_{n_{1}}\left(  \frac{\left(  a_{1}^{2}+a_{3}^{2}\right)  ^{2}%
		+a_{2}^{4}}{\left(  a_{1}^{2}+a_{3}^{2}\right)  ^{2}-a_{2}^{4}}\right)  \\
	 =&\left(  \left(  \cos\theta\cos\phi\right)  ^{2}+\left(  \cos\theta\sin
	\phi\sin\varphi-\sin\theta\cos\varphi\right)  ^{2}-\left(  \sin\theta
	\sin\varphi+\cos\theta\cos\varphi\sin\phi\right)  ^{2}\right)  ^{n_{3}}\\
	&  {\mathcal P}_{n_{1}}\left(  \frac{\left(  \left(  \cos\theta\cos\phi\right)
		^{2}+\left(  \cos\theta\sin\phi\sin\varphi-\sin\theta\cos\varphi\right)
		^{2}\right)  ^{2}+\left(  \sin\theta\sin\varphi+\cos\theta\cos\varphi\sin
		\phi\right)  ^{4}}{\left(  \cos\theta\cos\phi\right)  ^{2}+\left(  \cos
		\theta\sin\phi\sin\varphi-\sin\theta\cos\varphi\right)  ^{2}-\left(
		\sin\theta\sin\varphi+\cos\theta\cos\varphi\sin\phi\right)  ^{2}}\right)
		\nonumber
\end{align}
displays the results for $n_1=1,2$
\begin{align}
P_{1,0,0}^{B}\left(  \theta,\varphi,\phi\right)  =&\left(  \left(  \cos
\theta\cos\phi\right)  ^{2}+\left(  \cos\theta\sin\phi\sin\varphi-\sin
\theta\cos\varphi\right)  ^{2}\right)  ^{2}+\left(  \sin\theta\sin\varphi
+\cos\theta\cos\varphi\sin\phi\right)  ^{4}
\\
	P_{2,0,0}^{B}\left(  \theta,\varphi,\phi\right)     =&\frac{3}{2}\left(
	\left(  \left(  \cos\theta\cos\phi\right)  ^{2}+\left(  \cos\theta\sin\phi
	\sin\varphi-\sin\theta\cos\varphi\right)  ^{2}\right)  ^{2}+\left(  \sin
	\theta\sin\varphi+\cos\theta\cos\varphi\sin\phi\right)  ^{4}\right)  ^{2}\nonumber\\
	& -\frac{1}{2}\left(  \left(  \cos\theta\cos\phi\right)  ^{2}+\left(
	\cos\theta\sin\phi\sin\varphi-\sin\theta\cos\varphi\right)  ^{2}-\left(
	\sin\theta\sin\varphi+\cos\theta\cos\varphi\sin\phi\right)  ^{2}\right)  ^{2}.
\end{align}

\subsection{C-AB interaction}
To provide a comprehensive overview of all possible configurations, we will examine the \(\textbf{C-AB}\) interaction in the context of the Schmidt decomposition given by \eqref{546}
\begin{align}
	\psi_{n_{1},n_{2},n_{3}}^{ABC}\left(  x_{1},x_{2},x_{3}\right)    
{\textstyle\sum\limits_{l=0}^{n_{1}+n_{2}+n_{3}}}
\sqrt{\beta_{l}}\ \phi_{l}^{B}\left(  x_{2}\right)\  \Phi_{l}^{AC}\left(
x_{1},x_{3}\right) \label{545}.	 
\end{align}
Again, we will look at the three different scenarios: \((n_{1}=0, n_{2}=0, n_{3}\neq 0)\), \((n_{1}=0, n_{2}\neq 0, n_{3}=0)\), and \(( n_{1}\neq 0, n_{2}=0, n_{3}=0)\).

\subsubsection*{\bf D1. \ C-AB case 1: $n_{1}=0, n_{2}=0, n_{3}\neq 0$}

Substituting \(n_{1}=0, n_{2}=0\) into \eqref{gamma} and using \eqref{Afin} together with \eqref{335}, \eqref{338}, \eqref{331}, we get the following Schmidt modes
\begin{align}
\gamma_{\sigma}\equiv\gamma_{\sigma}^{0,0,n_{3}}&=
{\sum_{k=0}^{n_{3}-\sigma}}
\left(  A_{0,0,n_{3}}^{k,n_{3}-k-\sigma}\right)  ^{2}\\
=&\frac{n_{3}!\left(
	c_{3}^{2}\right)  ^{\sigma}}{\sigma!\left(  n_{3}-\sigma\right)  !}\left(
c_{2}^{2}+c_{1}^{2}\right)  ^{n_{3}-\sigma}%
\end{align}
verifying the condition
\begin{equation}
	\tr\rho_{0,0,n_{3}}^{C}=\tr\rho_{0,0,n_{3}}^{AB}=%
	{\sum_{\sigma=0}^{n_{3}}}
	\gamma_{\sigma}^{0,0,n_{3}}=\left(  c_{3}^{2}+c_{1}^{2}+c_{2}^{2}\right)
	^{n_{3}}=1.
\end{equation}
Then  the appropriate purity takes  the form
\begin{align}\label{PC00n3}
	P_{0,0,n_{3}}^{C}\left(  \theta,\varphi,\phi\right)     =&\left(  \left(
	c_{1}^{2}+c_{2}^{2}\right)  ^{2}-c_{3}^{4}\right)  ^{n_{3}}P_{n_{3}}\left(
	\frac{\left(  c_{1}^{2}+c_{2}^{2}\right)  ^{2}+c_{3}^{4}}{\left(  c_{1}%
		^{2}+c_{2}^{2}\right)  ^{2}-c_{3}^{4}}\right)  \\
	 =&\left(  \cos^{2}\phi\sin^{2}\theta+\left(  \cos\theta\sin\varphi-\sin
	\theta\cos\varphi\sin\phi\right)  ^{2}-\left(  \cos\theta\cos\varphi
	+\sin\theta\sin\phi\sin\varphi\right)  ^{2}\right)  ^{n_{3}}\\
	&  P_{n_{3}}\left(  \frac{\left(  \cos^{2}\phi\sin^{2}\theta+\left(
		\cos\theta\sin\varphi-\sin\theta\cos\varphi\sin\phi\right)  ^{2}\right)
		^{2}+\left(  \cos\theta\cos\varphi+\sin\theta\sin\phi\sin\varphi\right)  ^{4}%
	}{\cos^{2}\phi\sin^{2}\theta+\left(  \cos\theta\sin\varphi-\sin\theta
		\cos\varphi\sin\phi\right)  ^{2}-\left(  \cos\theta\cos\varphi+\sin\theta
		\sin\phi\sin\varphi\right)  ^{2}}\right)\nonumber
\end{align}

\subsubsection*{\bf D2. \  C-AB case 2: $n_{1}=0, n_{2}\neq0, n_{3}= 0$}

We analyze \eqref{gamma} under the conditions \( n_{1} = 0 \) and \( n_{3} = 0 \), along with \eqref{Afin}, as well as  \eqref{333}, \eqref{337}, and \eqref{330}, in order to derive the Schmidt modes
\begin{equation}
	\gamma_{\sigma}^{0,n_{2},0}=%
	{\sum_{k=0}^{n_{2}-\sigma}}
	\left(  A_{0,n_{2},0}^{k,n_{2}-k-\sigma}\right)  ^{2}=\tfrac{n_{2}!\left(
		b_{3}^{2}\right)  ^{\sigma}}{\sigma!\left(  n_{2}-\sigma\right)  !}\left(
	b_{1}^{2}+b_{2}^{2}\right)  ^{n_{2}-\sigma}.
\end{equation}
Using these quantities, it is easy to validate von Neumann's standard normalization condition
\begin{equation}
	\tr\rho_{0,n_{2},0}^{C}=\tr\rho_{0,n_{2},0}^{AB}=%
	{\sum_{\sigma=0}^{n_{2}}}
	\gamma_{\sigma}^{0,n_{2},0}=\left(  b_{1}^{2}+b_{3}^{2}+b_{2}^{2}\right)
	^{n_{2}}=1.
\end{equation}
As a result, we get the purity
\begin{equation}
	P_{n_{1},n_{2},n_{3}}^{C}\left(  \theta,\varphi,\phi\right)  =P_{n_{1}%
		,n_{2},n_{3}}^{AB}\left(  \theta,\varphi,\phi\right)  =\tr\left(  \rho
	_{n_{1},n_{2},n_{3}}^{C}\right)  ^{2}=
	{\textstyle\sum\limits_{\sigma=0}^{n_{1}+n_{2}+n_{3}}}
	\left(  \gamma_{\sigma}^{n_{1},n_{2},n_{3}}\right)  ^{2}%
\end{equation}
and especially for $\left( n_{1}=0,n_{2},n_{3}=0\right)$, we have the following
\begin{align}
	P_{0,n_{2},0}^{C}\left(  \theta,\varphi,\phi\right)    & =%
	{\textstyle\sum\limits_{\sigma=0}^{n_{2}}}
	\left(  \gamma_{\sigma}^{0,n_{2},0}\right)  ^{2}=\left(  \left(  b_{1}%
	^{2}+b_{2}^{2}\right)  ^{2}-b_{3}^{4}\right)  ^{n_{2}}P_{n_{2}}\left(
	\frac{\left(  b_{1}^{2}+b_{2}^{2}\right)  ^{2}+b_{3}^{4}}{\left(  b_{1}%
		^{2}+b_{2}^{2}\right)  ^{2}-b_{3}^{4}}\right)  \\
	& =\left(  \sin^{2}\phi+\cos^{2}\phi\cos^{2}\varphi-\cos^{2}\phi\sin
	^{2}\varphi\right)  ^{n_{2}}\times P_{n_{2}}\left(  \frac{\left(  \sin^{2}\phi+\cos^{2}\phi\cos
		^{2}\varphi\right)  ^{2}+\cos^{4}\phi\sin^{4}\varphi}{\sin^{2}\phi+\cos
		^{2}\phi\cos^{2}\varphi-\cos^{2}\phi\sin^{2}\varphi}\right)
\end{align}
For further illustration, for $n_2=1,2$ we find
\begin{align}
&P_{0,1,0}^{C}\left(  \theta,\varphi,\phi\right)  =\left(  \sin^{2}\phi
+\cos^{2}\phi\cos^{2}\varphi\right)  ^{2}+\cos^{4}\phi\sin^{4}\varphi
\\
&
	P_{0,2,0}^{C}\left(  \theta,\varphi,\phi\right)    =\frac{3}{2}\left(
	\left(  \sin^{2}\phi+\cos^{2}\phi\cos^{2}\varphi\right)  ^{2}+\cos^{4}\phi
	\sin^{4}\varphi\right)  ^{2} -\frac{1}{2}\left(  \sin^{2}\phi+\cos^{2}\phi\cos^{2}\varphi-\cos^{2}%
	\phi\sin^{2}\varphi\right)  ^{2}.
\end{align}%

\subsubsection*{\bf D3. \ C-AB case 3: $n_{1}\neq 0, n_{2}=0, n_{3}= 0$}

In the last scenario, starting from \eqref{beta} with \( n_{2} = 0 \) and \( n_{3} = 0 \), and using \eqref{Afin} along with equations \eqref{333}, \eqref{336}, and \eqref{339}, we arrive at the Schmidt modes	
	\begin{align}
		\gamma_{\sigma}^{n_{1},0,0}=&%
		{\sum_{l=0}^{n_{1}-\sigma}}
		\left(  A_{n_{1},0,0}^{k,n_{1}-k-\sigma}\right)  ^{2}\\
		=&\tfrac{n_{1}!\left(
			a_{3}^{2}\right)  ^{\sigma}}{\sigma!\left(  n_{1}-\sigma\right)  !}\left(
		a_{1}^{2}+a_{2}^{2}\right)  ^{n_{1}-\sigma}%
	\end{align}
and we have the condition%
	\begin{equation}
		\tr\rho_{n_{1},0,0}^{C}=\tr\rho_{n_{1},0,0}^{AB}=%
		{\sum_{k=0k}^{n_{1}\alpha n_{1},0,0}}
		=\left(  a_{1}^{2}+a_{3}^{2}+a_{2}^{2}\right)  ^{n_{1}}=1
	\end{equation}
	We show that the corresponding  purity is given by
	\begin{align}
		P_{n_{1},0,0}^{C}\left(  \theta,\varphi,\phi\right)     =&
		\left(  \left(  a_{1}%
		^{2}+a_{2}^{2}\right)  ^{2}-a_{3}^{4}\right)  ^{n_{1}}P_{n_{1}}\left(
		\frac{\left(  a_{1}^{2}+a_{2}^{2}\right)  ^{2}+a_{3}^{4}}{\left(  a_{1}%
			^{2}+a_{2}^{2}\right)  ^{2}-a_{3}^{4}}\right)  \\
		 =&\left(  \cos^{2}\theta\cos^{2}\phi+\left(  \sin\theta\sin\varphi+\cos
		\theta\cos\varphi\sin\phi\right)  ^{2}-\left(  \cos\theta\sin\phi\sin
		\varphi-\sin\theta\cos\varphi\right)  ^{2}\right)  ^{n_{1}}\nonumber\\
		&  P_{n_{1}}\left(  \tfrac{\left(  \cos^{2}\theta\cos^{2}\phi+\left(
			\sin\theta\sin\varphi+\cos\theta\cos\varphi\sin\phi\right)  ^{2}\right)
			^{2}+\left(  \cos\theta\sin\phi\sin\varphi-\sin\theta\cos\varphi\right)  ^{4}%
		}{\cos^{2}\theta\cos^{2}\phi+\left(  \sin\theta\sin\varphi+\cos\theta
			\cos\varphi\sin\phi\right)  ^{2}-\left(  \cos\theta\sin\phi\sin\varphi
			-\sin\theta\cos\varphi\right)  ^{2}}\right)
	\end{align}
which results in the following for $n_1=1,2$
\begin{align}
	P_{1,0,0}^{C}\left(  \theta,\varphi,\phi\right)  =&\left(  \cos^{2}\theta
	\cos^{2}\phi+\left(  \sin\theta\sin\varphi+\cos\theta\cos\varphi\sin
	\phi\right)  ^{2}\right)  ^{2}+\left(  \cos\theta\sin\phi\sin\varphi
	-\sin\theta\cos\varphi\right)  ^{4}%
	\\
	P_{2,0,0}^{C}\left(  \theta,\varphi,\phi\right)    = &\frac{3}{2}\left(
		\left(  \cos^{2}\theta\cos^{2}\phi+\left(  \sin\theta\sin\varphi+\cos
		\theta\cos\varphi\sin\phi\right)  ^{2}\right)  ^{2}+\left(  \cos\theta\sin
		\phi\sin\varphi-\sin\theta\cos\varphi\right)  ^{4}\right)  ^{2}\nonumber\\
		& -\frac{1}{2}\left(  \cos^{2}\theta\cos^{2}\phi+\left(  \sin\theta\sin
		\varphi+\cos\theta\cos\varphi\sin\phi\right)  ^{2}-\left(  \cos\theta\sin
		\phi\sin\varphi-\sin\theta\cos\varphi\right)  ^{2}\right)  ^{2}.
	\end{align}

\subsection{Illustration and numerical examples}
In order to highlight the main differences between the various interactions and to provide a clearer understanding of the cases we have considered so far, we will focus on the equations \eqref{PA00n3}, \eqref{PB00n3}, and \eqref{PC00n3}. These correspond to the configuration where the quantum numbers are defined as \( n_1 = 0 \), \( n_2 = 0 \), and \( n_3 \neq 0 \). We will set \( n_3 = 1, 2 \) for further analysis. 

\subsection*{{\bf E1. \ Purity function $	P_{0,0,n_3}^{A}$}}
\begin{figure}[H]
		\centering
		{\includegraphics[width=0.45\linewidth]{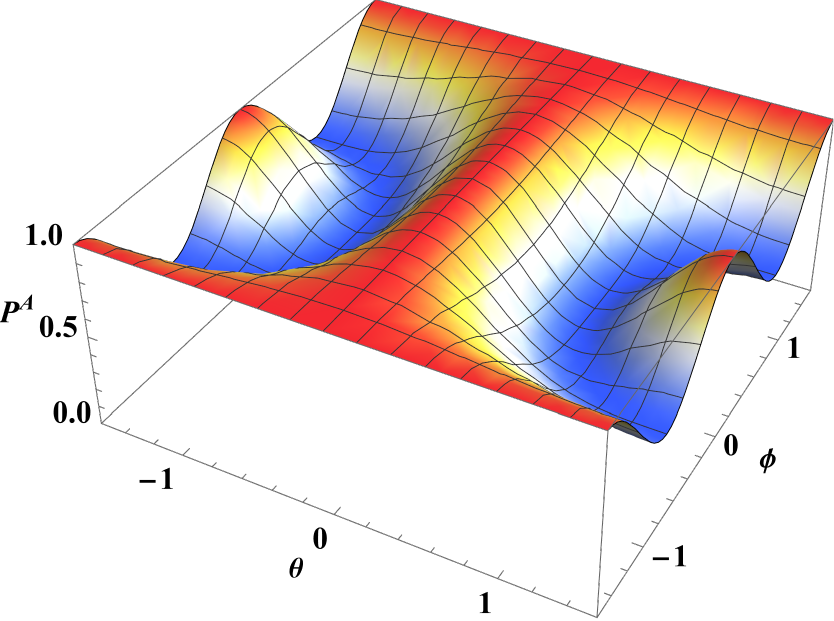}}
		{\caption{(Color online) The purity function $P^A_{0,0,1}=P_a$ versus the two angles $\theta$ and $\phi$ $\in[-\pi, \pi]$. 
			}\label{pa1}}
	\end{figure}

As far as \eqref{PA00n3} is concerned, then for \( n_3 = 1, 2 \) we get
\begin{align}
&	P_{0,0,1}^{A}\left(  \theta,\varphi,\phi\right)  
	=\left(  \cos^{2}\theta+\sin^{2}\theta\sin^{2}%
	\phi\right)  ^{2}+\sin^{4}\theta\cos^{4}\phi\label{P001}\\
&	P_{0,0,2}^{A}\left(  \theta,\varphi,\phi\right)    
	=\frac{3}{2}\left(  \left(  \cos^{2}\theta+\sin^{2}\theta\sin
	^{2}\phi\right)  ^{2}+\sin^{4}\theta\cos^{4}\phi\right)  ^{2}-\frac{1}{2}\left(  \left(
	\cos^{2}\theta+\sin^{2}\theta\sin^{2}\phi\right)  ^{2}-\sin^{4}\theta\cos
	^{4}\phi\right)  ^{2}\label{P002}
\end{align}
In Fig. \ref{pa1}, we plot the purity function $ P_{0,0,1}^{A}$ \eqref{P001}  as functions of the wo angles $\theta$ and $\phi$.
We observe several peaks and valleys, indicating significant variation in purity as a function of angle. The peaks correspond to higher purity values, approaching 1, while the valleys  represent lower purity values, around 0.5. These variations suggest that certain angular configurations lead to a more pure (less mixed) quantum state, while others lead to a more mixed state. This sensitivity of purity to changes in the angles \( \theta \) and \( \phi \) may reflect how the quantum state evolves under transformations parameterized by these angles. The surface also exhibits periodicity, a common feature in functions that depend on angular variables, which may be related to symmetries in the underlying quantum system, such as rotational symmetry. The pattern of peaks and troughs likely reflects the intrinsic symmetries of the quantum state, possibly related to rotational invariance or the specific symmetries of the Hamiltonian governing the system.

\subsection*{{\bf E2. \ Purity function $	P_{0,0,n_3}^{B}$}}	
Regarding  \eqref{PB00n3} with	 $n_3=1,2$, we find 
	\begin{align}
		P_{0,0,1}^{B}\left(  \theta,\varphi,\phi\right)     =&\left(  \cos^{2}\phi
		\sin^{2}\theta+\left(  \cos\theta\cos\varphi+\sin\theta\sin\phi\sin
		\varphi\right)  ^{2}\right)  ^{2}
		+\left(  \cos\theta\sin\varphi-\sin\theta\cos\varphi\sin\phi\right)  ^{4} \label{pb001}
		\\
		P_{0,0,2}^{B}\left(  \theta,\varphi,\phi\right)    =& \frac{3}{2}\left(
		\left(  \cos^{2}\phi\sin^{2}\theta+\left(  \cos\theta\cos\varphi+\sin
		\theta\sin\phi\sin\varphi\right)  ^{2}\right)  ^{2}+\left(  \cos\theta
		\sin\varphi-\sin\theta\cos\varphi\sin\phi\right)  ^{4}\right)  ^{2}\nonumber\\
		& -\frac{1}{2}\left(  \cos^{2}\phi\sin^{2}\theta+\left(  \cos\theta\cos
		\varphi+\sin\theta\sin\phi\sin\varphi\right)  ^{2}-\cos^{2}\theta\sin
		^{2}\varphi\right)  ^{2}%
	\end{align}%
	
\begin{figure}[H]
		\centering
		{\includegraphics[width=0.45\linewidth]{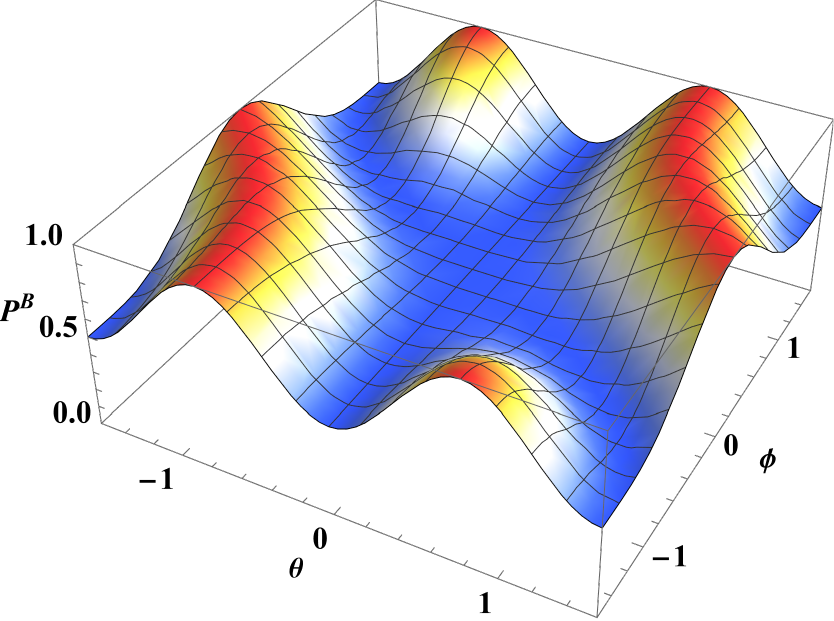}}
		{\caption{(Color online) The purity function $P^B_{0,0,1}=P_b$ versus the two angles $\theta$ and $\phi$ $\in[-\pi, \pi]$ with $\varphi=\frac{\pi}{2}$. 
			}\label{pb1}}
	\end{figure}

	Fig. \ref{pb1} shows the purity $P_{0,0,1}^{B}$ \eqref{pb001} as a function of the two angles $\theta$ and $\phi$ $\in[-\pi, \pi]$ with $\varphi=\frac{\pi}{4}$. When this purity is consistently lower than in Fig. \ref{pa1}, this may indicate that the entanglement dynamics are less favorable for this particular interaction. We note that any oscillations in purity could reflect coherent oscillations between the oscillators, indicating a periodic exchange of entanglement. A notable increase in purity at certain coupling strengths may indicate resonant interactions or optimal conditions for entanglement generation.

	\subsection*{{\bf E3. \ Purity function $	P_{0,0,n_3}^{C}$}}
By  choosing $n_3=1,2$, \eqref{PC00n3} gives
	\begin{align}
		P_{0,0,1}^{C}\left(  \theta,\varphi,\phi\right)  =&\left(  \left(  \cos
		\theta\sin\varphi-\sin\theta\cos\varphi\sin\phi\right)  ^{2}+\cos^{2}\phi
		\sin^{2}\theta\right)  ^{2}+\left(  \cos\theta\cos\varphi+\sin\theta\sin
		\phi\sin\varphi\right)  ^{4}\label{pc001}
		\\
		P_{0,0,2}^{C}\left(  \theta,\varphi,\phi\right)     =&\frac{3}{2}\left(
		\left(  \left(  \cos\theta\sin\varphi-\sin\theta\cos\varphi\sin\phi\right)
		^{2}+\cos^{2}\phi\sin^{2}\theta\right)  ^{2}+\left(  \cos\theta\cos
		\varphi+\sin\theta\sin\phi\sin\varphi\right)  ^{4}\right)  ^{2}\nonumber\\
		& -\frac{1}{2}\left(  \left(  \cos\theta\sin\varphi-\sin\theta\cos\varphi
		\sin\phi\right)  ^{2}+\cos^{2}\phi\sin^{2}\theta-\left(  \cos\theta\cos
		\varphi+\sin\theta\sin\phi\sin\varphi\right)  ^{2}\right)  ^{2}%
	\end{align}%

\begin{figure}[H]
		\centering
		{\includegraphics[width=0.45\linewidth]{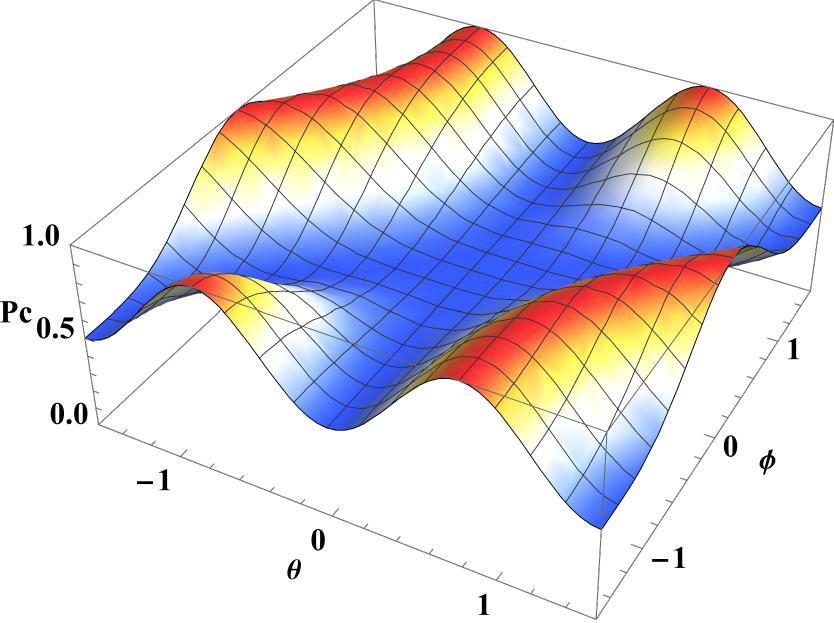}}
		{\caption{(Color online) The purity function $P^C_{0,0,1}=P_c$ versus the two angles $\theta$ and $\phi$ $\in[-\pi, \pi]$ with $\varphi=\frac{\pi}{4}$. }\label{pc1}}
	\end{figure}

	Fig \ref{pc1} shows the purity $P_{0,0,1}^{C}$ \eqref{pc001} as function of the two angles $\theta$ and $\phi$ $\in[-\pi, \pi]$ with $\varphi=\frac{\pi}{4}$. If this interaction has the highest purity of the three, it would indicate that {\bf C-AB} is particularly well entangled with {\bf A-BC} and  {\bf B-AC}, suggesting effective coupling.  A gradual increase or consistent purity could indicate stable entanglement over a range of parameters, highlighting the resilience of this particular interaction. Implications for quantum systems: The results may imply that the CC subsystem plays a crucial role in maintaining or enhancing entanglement in the system, which could be useful for quantum information applications.

	By comparing the three figures, we can draw conclusions about the effectiveness of each interaction in generating entanglement. Certain interactions may be more advantageous under certain conditions, while others may show instability or lower entanglement purity. Together, the figures illustrate the dynamic interplay between the subsystems and highlight how different coupling scenarios can lead to different entanglement outcomes. These interpretations may guide further investigations to optimize coupling configurations for enhanced entanglement in practical quantum systems.
	
\section{Conclution}

We developed a method to analytically solve for the energy eigenstates of three coupled quantum harmonic oscillators, expressing the tripartite wavefunction in a separable form. By applying the Schmidt decomposition, we represented the wavefunction as a sum of orthogonal product states describing bipartitions. By integrating against the Schmidt modes, we obtained closed-form Schmidt coefficients that quantify the bipartite entanglement. In particular, we showed that these coefficients are encoded in a hypergeometric function, providing a compact characterization of the reduced density matrices. 
Tuning the interaction strengths affected the dynamics of the coefficients, and comparing them revealed the distribution of entanglement across bipartitions. As the couplings varied, we gained insight into multipartite sharing within a three coupled harmonic oscillators. This work established a framework for analyzing  static  bipartition entanglement in this model.

In summary, we have developed a substantial solution for the characterization of tripartite entanglement, providing deeper insights into the entanglement structure of quantum systems with three coupled subsystems. This result not only advances our understanding of multipartite entanglement, but also has broader implications for a variety of quantum systems, such as quantum information processing, quantum computing, and entanglement-based technologies. Our approach paves the way for more sophisticated control and manipulation of entangled states, which is critical for advancing both theoretical and practical applications in quantum science.

Finally, we emphasize that here we have applied the idea introduced by Makarov \cite{Makarov} regarding the computation of Schmidt numbers using Schmidt decomposition to a non-Gaussian bipartite system. This approach allowed us to gain valuable insights into the entanglement properties of such systems. Building on this foundation, we extended the methodology to a non-Gaussian tripartite system, which presents additional complexities and opportunities for exploration. In our study, we focused on a specific class of tripartite systems to provide a detailed analysis of their properties and behavior. This focused approach allowed us to highlight key aspects of entanglement and information distribution in these systems.We recognize that a thorough classification of various tripartite configurations and their respective Schmidt numbers is a significant undertaking. Therefore, we have reserved a comprehensive classification analysis for a future work in which we aim to systematically explore the broader implications of our findings for different classes of non-Gaussian tripartite systems. This future work will build on our current results and provide a more complete understanding of the entanglement structures at play.

\appendix
\section{Energy spectrum}\label{EESS}

Using an algebraic method based on the Lie group $SU(3)$, we showed that the eigenstates  are given by the following wave functions in terms of the Hermite polynomials \cite{MerdaciPLA2020}
\begin{align}
	\psi_{n_{1},n_{2},n_{3}}\left(  x_{1},x_{2},x_{3}\right)    =&\left(
	\frac{m\varpi}{\pi\hbar}\right)  ^{\frac{3}{4}}\frac{1}{\sqrt{2^{n_{1}%
				+n_{2}+n_{3}}n_{1}!n_{2}!n_{3}!}}e^{-\tfrac{m\varpi}{2\hbar}\left(
		e^{\varsigma-\rho}q_{1}^{2}+e^{\kappa-\varsigma}q_{2}^{2}+e^{\rho-\kappa}%
		q_{3}^{2}\right)  }\nonumber\label{states}\\
	&   H_{n_{1}}\left(  \sqrt{\tfrac{m\varpi e^{\varsigma-\rho}}{\hbar}%
	}q_{1}\right)  H_{n_{2}}\left(  \sqrt{\tfrac{m\varpi e^{\kappa-\varsigma}%
		}{\hbar}}q_{2}\right)  H_{n_{3}}\left(  \sqrt{\tfrac{m\varpi e^{\rho-\kappa}%
		}{\hbar}}q_{3}\right).
\end{align}
The associated eigenvalues have the form
\begin{equation}
	E_{n_{1},n_{2},n_{3}}=\hbar\varpi\left(  e^{\varsigma-\rho}n_{1}%
	+e^{\kappa-\varsigma}n_{2}+e^{\rho-\kappa}n_{3}+\frac{e^{\varsigma-\rho
		}+e^{\kappa-\varsigma}+e^{\rho-\kappa}}{2}\right)
\end{equation}
where the new variables are mapped to the old as
\begin{align}
	q_{1}  &  =\mu_{1}\cos\theta\cos\phi x_{1}-\mu_{2}\left(  \sin\theta
	\sin\varphi+\cos\theta\cos\varphi\sin\phi\right)  x_{2}-\mu_{3}\left(
	\sin\theta\cos\varphi-\cos\theta\sin\phi\sin\varphi\right)  x_{3}\label{422}\\
	q_{2}  &  =\mu_{1}\sin\phi x_{1}+\mu_{2}\cos\phi\cos\varphi x_{2}-\mu_{3}%
	\cos\phi\sin\varphi x_{3}\label{423}\\
	q_{3}  &  =\mu_{1}\cos\phi\sin\theta x_{1}+\mu_{2}\left(  \cos\theta
	\sin\varphi-\sin\theta\cos\varphi\sin\phi\right)  x_{2}+\mu_{3}\left(
	\cos\theta\cos\varphi+\sin\theta\sin\phi\sin\varphi\right)  x_{3}. \label{424}%
\end{align} 
and we have set the following quantities  
\begin{align}
	&
	m=\left(m_1 m_2 m_3\right)^{\frac{1}{3}}, \quad \mu_i=\left(\frac{m_i}{m}\right)^{\frac{1}{2}}, \quad \mu_1 \mu_2 \mu_3=1, \quad J_{i j}=\frac{D_{i j}}{2 \sqrt{m_i m_j}}
	\\
	&
	\varpi=\left(  \Sigma_{1}\Sigma_{2}\Sigma_{3}\right)
	^{\frac{1}{3}}, \quad e^{\varsigma-\rho}=\frac{\Sigma_{1}}{\varpi},\quad
	e^{\kappa-\varsigma}=\frac{\Sigma_{2}}{\varpi},\quad e^{\rho-\kappa}%
	=\frac{\Sigma_{3}}{\varpi}.
\end{align}

The old and new parameters are related as follows
\begin{align}
	\omega_{1}^{2}  &  =\left(  \Sigma_{1}^{2}\cos^{2}\theta+\Sigma_{3}^{2}%
	\sin^{2}\theta\right)  \cos^{2}\phi+\Sigma_{2}^{2}\sin^{2}\phi\label{o11}\\
	\omega_{2}^{2}  &  =2\tfrac{\left(  \Sigma_{2}^{2}\cos^{2}\phi+\left(
		\Sigma_{1}^{2}\cos^{2}\theta+\Sigma_{3}^{2}\sin^{2}\theta\right)  \sin^{2}%
		\phi\right)  \cos^{2}\varphi+\left(  \Sigma_{3}^{2}\cos^{2}\theta+\Sigma
		_{1}^{2}\sin^{2}\theta\right)  \sin^{2}\varphi}{2}+\tfrac{\Sigma_{1}%
		^{2}-\Sigma_{3}^{2}}{2}\sin2\theta\sin\phi\sin2\varphi\label{o22}\\
	\omega_{3}^{2}  &  =2\tfrac{\left(  \Sigma_{3}^{2}\cos^{2}\theta+\Sigma
		_{1}^{2}\sin^{2}\theta\right)  \cos^{2}\varphi+\left(  \Sigma_{2}^{2}\cos
		^{2}\phi+\left(  \Sigma_{1}^{2}\cos^{2}\theta+\Sigma_{3}^{2}\sin^{2}%
		\theta\right)  \sin^{2}\phi\right)  \sin^{2}\varphi}{2}-\tfrac{\Sigma_{1}%
		^{2}-\Sigma_{3}^{2}}{2}\sin2\theta\sin\phi\sin2\varphi\label{o33}\\
	J_{12}  &  =-\tfrac{\left(  \Sigma_{1}^{2}-\Sigma_{2}^{2}\right)  \cos
		^{2}\theta+\left(  \Sigma_{3}^{2}-\Sigma_{2}^{2}\right)  \sin^{2}\theta}%
	{2}\sin2\phi\cos\varphi-\tfrac{\Sigma_{1}^{2}-\Sigma_{3}^{2}}{2}\sin
	2\theta\cos\phi\sin\varphi\label{j12}\\
	J_{13}  &  =\tfrac{\left(  \Sigma_{1}^{2}-\Sigma_{2}^{2}\right)  \cos
		^{2}\theta+\left(  \Sigma_{3}^{2}-\Sigma_{2}^{2}\right)  \sin^{2}\theta}%
	{2}\sin2\phi\sin\varphi-\tfrac{\Sigma_{1}^{2}-\Sigma_{3}^{2}}{2}\sin
	2\theta\cos\phi\cos\varphi\label{j13}\\
	J_{23}  &  =\tfrac{\Sigma_{1}^{2}-\Sigma_{3}^{2}}{2}\sin2\theta\sin\phi
	\cos2\varphi-\tfrac{\left(  \left(  \Sigma_{2}^{2}-\Sigma_{3}^{2}\right)
		\cos^{2}\theta-\left(  \Sigma_{1}^{2}-\Sigma_{2}^{2}\right)  \sin^{2}%
		\theta\right)  \cos^{2}\phi+\left(  \Sigma_{1}^{2}-\Sigma_{3}^{2}\right)
		\cos2\theta\sin^{2}\phi}{2}\sin2\varphi\label{j23}.
\end{align}
The ratios between the coupling parameters \( J_{ij} \) ($i, j=1,2,3$) and the squared frequency differences \( \omega_{i}^{2} - \omega_{j}^{2} \) can be readily determined. They are expressed as:
\begin{align}
	\tfrac{2J_{12}}{\omega_{1}^{2}-\omega_{2}^{2}}  &  =\tfrac{-\left(  \Sigma
		_{1}^{2}-\Sigma_{2}^{2}\right)  \cos^{2}\theta\sin2\phi\cos\varphi+\left(
		\Sigma_{2}^{2}-\Sigma_{3}^{2}\right)  \sin^{2}\theta\sin2\phi\cos
		\varphi+\left(  \Sigma_{3}^{2}-\Sigma_{1}^{2}\right)  \sin2\theta\cos\phi
		\sin\varphi}{\left(  \Sigma_{2}^{2}-\Sigma_{3}^{2}\right)  \left(  \sin
		^{2}\phi-\cos^{2}\phi\cos^{2}\varphi\right)  +\left(  \Sigma_{3}^{2}%
		-\Sigma_{1}^{2}\right)  \left(  \sin^{2}\theta\sin^{2}\varphi-\cos^{2}%
		\theta\cos^{2}\phi+\cos^{2}\theta\sin^{2}\phi\cos^{2}\varphi+\tfrac{1}{2}%
		\sin2\theta\sin\phi\sin2\varphi\right)  }\label{A14}\\
	\tfrac{2J_{13}}{\omega_{1}^{2}-\omega_{3}^{2}}  &  =\tfrac{\left(  \Sigma
		_{1}^{2}-\Sigma_{2}^{2}\right)  \cos^{2}\theta\sin2\phi\sin\varphi-\left(
		\Sigma_{2}^{2}-\Sigma_{3}^{2}\right)  \sin^{2}\theta\sin2\phi\sin
		\varphi+\left(  \Sigma_{3}^{2}-\Sigma_{1}^{2}\right)  \sin2\theta\cos\phi
		\cos\varphi}{\left(  \Sigma_{3}^{2}-\Sigma_{1}^{2}\right)  \left(  \sin
		^{2}\theta\cos^{2}\varphi-\cos^{2}\theta\cos^{2}\phi+\cos^{2}\theta\sin
		^{2}\phi\sin^{2}\varphi-\tfrac{1}{2}\sin2\theta\sin\phi\sin2\varphi\right)
		+\left(  \Sigma_{2}^{2}-\Sigma_{3}^{2}\right)  \left(  \sin^{2}\phi-\cos
		^{2}\phi\sin^{2}\varphi\right)  }\label{A15}\\
	\tfrac{2J_{23}}{\omega_{2}^{2}-\omega_{3}^{2}}  &  =\tfrac{\left(  \Sigma
		_{3}^{2}-\Sigma_{1}^{2}\right)  \left(  \cos2\theta\sin^{2}\phi\sin
		2\varphi-\sin2\theta\sin\phi\cos2\varphi\right)  -\left(  \Sigma_{2}%
		^{2}-\Sigma_{3}^{2}\right)  \cos^{2}\theta\cos^{2}\phi\sin2\varphi+\left(
		\Sigma_{1}^{2}-\Sigma_{2}^{2}\right)  \sin^{2}\theta\cos^{2}\phi\sin2\varphi
	}{\left(  \Sigma_{2}^{2}-\Sigma_{3}^{2}\right)  \cos^{2}\phi\cos
		2\varphi+\left(  \Sigma_{3}^{2}-\Sigma_{1}^{2}\right)  \left(  \sin^{2}%
		\theta\cos2\varphi-\cos^{2}\theta\sin^{2}\phi\cos2\varphi-\sin2\theta\sin
		\phi\sin2\varphi\right)  }\label{A16}.
\end{align}

\section{Particular  cases}\label{PPCC}


Let us examine the derivation of certain results already established in the literature, focusing on three different limit cases that are distinguished by the coupling parameters. To obtain solutions for two coupled harmonic oscillators in the variables $\left( x_{1},x_{2}\right)$, we only need to consider the limit $D_{13},D_{23}\longrightarrow0$, which implies $J_{13},J_{23}\longrightarrow0$. These operations constrain the Hamiltonian \eqref{Ham1} to the following form:
\begin{equation}
H= H_{0}+\frac{p_{3}^{2}}{2m_{3}}+\frac{1}{2}m_{3}\omega
_{3}^{2}x_{3}^{2}%
\end{equation}
where $H_{0}$ is the Hamiltonian of the two coupled harmonic oscillators in
$\left(  x_{1},x_{2}\right)  $ variables
\begin{equation}
H_{0}=\frac{p_{1}^{2}}{2m_{1}}+\frac{p_{2}^{2}}{2m_{2}}+\frac{1}{2}m_{1}%
\omega_{1}^{2}x_{1}^{2}+\frac{1}{2}m_{2}\omega_{2}^{2}x_{2}^{2}+\frac{1}%
{2}D_{12}x_{1}x_{2}.
\end{equation}
If we set $\tilde{J}_{13},\tilde{J}_{23}\longrightarrow0$ in (\ref{200}-\ref{211}), we find that $\theta\longrightarrow0$ and $\varphi\longrightarrow0$. As a result, \eqref{200} simplifies to the following relation
\begin{equation}
\tfrac{2J_{12}}{\omega_{1}^{2}-\omega_{2}^{2}}\longrightarrow\tfrac{2\tilde
{J}_{12}}{\tilde{\omega}_{1}^{2}-\tilde{\omega}_{2}^{2}}=-\tan2\phi.
\end{equation}
It is worth noting that the mentioned conditions correspond to those used in our previous study \cite{Jellal11} to decouple the problem of two harmonic oscillators in the variables $\left(x_{1}, x_{2}\right)$. The result is the transformed Hamiltonian 
\begin{equation}
H=\left(  \frac{p_{1}^{2}}{2m}+\frac{p_{2}^{2}}{2m}+\frac{m}{2}kq_{1}%
^{2}+\frac{m}{2}kq_{2}^{2}\right)  +\frac{p_{3}^{2}}{2m_{3}}+\frac{1}%
{2}m\omega_{3}^{2}q_{3}^{2}%
\end{equation}
with the variables
\begin{equation}
	q_{1}=\mu_{1}\cos\phi x_{1}-\mu_{2}\sin\phi x_{2},\quad q_{2}=\mu_{1}\sin\phi
	x_{1}+\mu_{2}\cos\phi x_{2},\quad q_{3}=\mu_{3}x_{3}.
\end{equation}
In this case, the parameters (\ref{333}-\ref{339}) are reduced to
\begin{align}
a_{1} &  =\cos\phi,\qquad b_{1}=\sin\phi,\qquad c_{1}=0\\
a_{2} &  =-\sin\phi,\qquad b_{2}=\cos\phi,\qquad c_{2}=0\\
a_{3} &  =0,\qquad b_{3}=0,\qquad c_{3}=1
\end{align}
which lead to the following four conditions that are fulfilled by the quantum numbers
\begin{equation}
i_{2}=k-i_{1},\quad j_{1}=n_{1}-i_{1},\quad j_{2}=n_{2}-k+i_{1},\quad
l=n_{1}+n_{2}-k.
\end{equation}
These can be substituted into \eqref{aa} to get the Schmidt modes
\begin{align}
	A_{n_{1},n_{2},n_{3}
}^{k,l}=A_{n_{1},n_{2}}^{k}\delta_{l,n_{1}+n_{2}-k}
\end{align} 
where $A_{n_{1},n_{2}}^{k}$ is
\begin{align}
A_{n_{1},n_{2}}^{k}
=\tfrac{\sqrt{k!\left(  n_{1}+n_{2}-k\right)  !}\left(  -\sin\phi\right)
^{n_{1}-k}\left(  \allowbreak\cos\phi\right)  ^{n_{2}-k}P_{k}^{\left(
n_{1}-k,n_{2}-k\right)  }\left(  \cos2\phi\right)  }{\sqrt{n_{1}!n_{2}!}}.
\end{align}

On the other hand, to make a connection between our results and those obtained by Makarov \cite{Makarov}, it is enough to use the two relations \cite{Derivatives}
\begin{align}
&P_{n}^{\left(  \rho,m-n\right)  }\left(  z\right)  =\tfrac{m!}{n!}%
\tfrac{\Gamma\left(  n+\rho+1\right)  }{\Gamma\left(  m+\rho+1\right)
}\left(  \tfrac{z+1}{2}\right)  ^{n-m}P_{m}^{\left(  \rho,n-m\right)  }\left(
z\right)
\\
&P_{n}^{\left(  \rho,\sigma\right)  }\left(  z\right)  =\left(  \tfrac{1-z}%
{2}\right)  ^{n}P_{n}^{\left(  -\rho-\sigma-2n-1,\sigma\right)  }\left(
\frac{z+3}{z-1}\right).
\end{align}
As a result, we show that
\begin{align}
\lambda_{k} &=\sum\limits_{l}\left(  A_{n_{1},n_{2},n_{3}}^{k,l}\right)
^{2}=\left(  A_{n_{1},n_{2}}^{k}\right)  ^{2}\\
&=\tfrac{n_{1}!n_{2}%
!\sin^{2\left(  n_{1}+k\right)  }\left(  \phi\right)  \cos^{2\left(
n_{2}-k\right)  }\left(  \phi\right)  \left(  P_{n_{1}}^{\left(  -n_{1}%
-n_{2}-1,n_{2}-k\right)  }\left(  -\frac{2+\tan^{2}\left(  \phi\right)  }%
{\tan^{2}\left(  \phi\right)  }\right)  \right)  ^{2}}{k!\left(  n_{1}%
+n_{2}-k\right)  !}.
\end{align}
In this case, the state \eqref{544} becomes%
\begin{equation}
\sum\limits_{l}A_{n_{1},n_{2},n_{3}}^{k,l}\varphi_{k}\left(  x_{1}\right)
\phi_{l}\left(  x_{2}\right)  \chi_{n_{1}+n_{2}+n_{3}-k-l}\left(
x_{3}\right)  \longrightarrow A_{n_{1},n_{2}}^{k}\varphi_{k}\left(
x_{1}\right)  \phi_{n_{1}+n_{2}-k}\left(  x_{2}\right)  \chi_{n_{3}}\left(
x_{3}\right)
\end{equation}
and the decomposition  is reduced to
\begin{equation}
\psi^{ABC}=\sum_{k}A_{n_{1},n_{2}}^{k}\varphi_{k}\left(  x_{1}\right)
\phi_{n_{1}+n_{2}-k}\left(  x_{2}\right)  \chi_{n_{3}}\left(  x_{3}\right)
=\psi^{AB}\left(  x_{1},x_{2}\right)  \chi_{n_{3}}\left(  x_{3}\right)
\end{equation}
where $\psi^{AB}\left(  x_{1},x_{2}\right) $  is
\begin{equation}
\psi^{AB}\left(  x_{1},x_{2}\right)  =\sum_{k}A_{n_{1},n_{2}}^{k}\varphi
_{k}\left(  x_{1}\right)  \phi_{n_{1}+n_{2}-k}\left(  x_{2}\right)
\end{equation}
which coincides exactly with the one obtained by Makarov \cite{Makarov}.


\end{document}